\documentclass[lettersize,journal]{IEEEtran}
\usepackage{amsmath,amsfonts}
\usepackage{algorithmic}
\usepackage[linesnumbered,ruled,vlined]{algorithm2e}
\usepackage{array}
\usepackage{textcomp}
\usepackage{stfloats}
\usepackage{url}
\usepackage{verbatim}
\usepackage{graphicx}
\usepackage{float}
\usepackage[caption=false,font=footnotesize,labelfont=rm,textfont=rm]{subfig}
\usepackage{cite}
\usepackage{booktabs}
\usepackage{threeparttable}
\usepackage{todonotes}
\usepackage{lipsum}
\usepackage{color}

\newtheorem{remark}{Remark}[section]
\newtheorem{assumption}{Assumption}[section]

\begin{document}

\title{Interaction-Aware Motion Planning for Autonomous Vehicles with Multi-Modal Obstacle Uncertainty Predictions}

\author{Jian Zhou,  Bj\"orn Olofsson, and Erik Frisk
\thanks{This research was supported by the Strategic Research Area at Link\"oping--Lund in Information Technology (ELLIIT). {\it (Corresponding author: Jian Zhou.)}}
\thanks{Jian Zhou and Erik Frisk are with the Division of Vehicular Systems, Department of Electrical Engineering, Link\"oping University, SE-581 83 Link\"oping, Sweden. E-mail:
	\texttt{\{jian.zhou, erik.frisk\}@liu.se}.}
\thanks{Bj\"orn Olofsson is with the Department of Automatic Control, Lund University, SE-221 00 Lund, Sweden, and also with the Division of Vehicular Systems, Department of Electrical Engineering, Link\"oping University, SE-581 83 Link\"oping, Sweden. E-mail:
	\texttt{bjorn.olofsson@control.lth.se}.}
}

\markboth{This paper has been accepted to publish by IEEE Transactions on Intelligent Vehicles (DOI: 10.1109/TIV.2023.3314709)}%
{Shell \MakeLowercase{\textit{et al.}}: A Sample Article Using IEEEtran.cls for IEEE Journals}


\maketitle

\begin{abstract}
This paper proposes an interaction and safety-aware motion-planning method for an autonomous vehicle in uncertain multi-vehicle traffic environments. The method integrates the ability of the interaction-aware interacting multiple model Kalman filter (IAIMM-KF) to predict interactive multi-modal maneuvers of surrounding vehicles, and the advantage of model predictive control (MPC) in planning an optimal trajectory in uncertain dynamic environments. The multi-modal prediction uncertainties, containing both the maneuver and trajectory uncertainties of surrounding vehicles, are considered in computing the reference targets and designing the collision-avoidance constraints of MPC for resilient motion planning of the ego vehicle. The MPC achieves safety awareness by incorporating a tunable parameter to adjust the predicted obstacle occupancy in the design of the safety constraints, allowing the approach to achieve a trade-off between performance and robustness. Based on the prediction of the surrounding vehicles, an optimal reference trajectory of the ego vehicle is computed by MPC to follow the time-varying reference targets and avoid collisions with obstacles. The efficiency of the method is illustrated in challenging highway-driving simulation scenarios and a driving scenario from a recorded traffic dataset.
\end{abstract}


\section{Introduction}\label{Introduction}
\IEEEPARstart{M}{otion} planning and control for autonomous vehicles are promising approaches for, among other things, increasing the safety of autonomous driving~\cite{olofsson2021using}. An autonomous ego vehicle (EV) predicts the motion of surrounding vehicles (SVs) and then uses the predictions in the motion-planning strategy to plan collision-free reference trajectories. One main complication is that, in dynamic multi-vehicle traffic environments, vehicle-to-vehicle interactions and prediction uncertainties of SVs are crucial for the proactivity and safety of the motion planner \cite{yoon2021interaction}, \cite{zhou2023interaction}. Therefore, a key research question is how to consider the interactions and uncertainties in the motion planner to realize interaction and safety-aware dynamic motion planning of the EV. 

Interaction-aware motion-prediction models, which are defined as the upper-level approach among the three prediction levels \cite{lefevre2014survey}, \cite{westny2021vehicle}, can predict the interactive behaviors of SVs. Recent research in \cite{zhou2022interaction} has verified that model predictive control (MPC) combined with an interaction-aware motion-prediction model, the interaction-aware interacting multiple model Kalman filter (IAIMM-KF) \cite{lefkopoulos2021interaction}, makes the motion-planning performance more proactive in multi-vehicle traffic scenarios. However, the motion of SVs cannot be exactly predicted. So, the inherent prediction uncertainties of SVs, which typically contain uni- or multi-modal motion uncertainties,  bring challenges to the robustness of the planner and finally affect the safety of autonomous driving \cite{brudigam2021collision}. 

Uni-modal uncertainty means, for example, that the EV is sure about which maneuver the SV will perform, while the specific trajectory to execute the maneuver is uncertain. Multi-modal uncertainties mean that the SV has multiple uncertain maneuvers, where the trajectory of each maneuver is also uncertain. Multi-modal uncertainties of SVs are more realistic in the real world, and thereby attracted a lot of attention in the literature. The developed interaction-aware MPC in \cite{zhou2022interaction} assumed that the SVs move as predicted by the EV, such that it did not have safety awareness in uncertain environments. In practice, systematic consideration of multi-modal uncertainties in an interaction-aware dynamic motion-planning strategy is an open problem \cite{chen2022interactive}, \cite{wang2022interaction}. To address these aspects, this paper proposes an interaction and safety-aware MPC (ISA-MPC) for dynamic motion planning in multi-vehicle traffic scenarios using the interaction-aware motion prediction of SVs, where both longitudinal and lateral multi-modal motion uncertainties of SVs are considered. The specific contributions of realizing and verifying the method are: 
\begin{enumerate}
	\item \textbf{Interaction and safety-aware motion planning}: The proposed ISA-MPC achieves interaction awareness based on the interaction-aware motion prediction by the IAIMM-KF, and safety awareness through tuning a safety-awareness parameter to model the uncertain obstacle occupancy. The tunable safety-awareness parameter empowers the method with the flexibility to achieve a trade-off between robustness and performance by adjusting the obstacle occupancy, while keeping the computational complexity of the algorithm unaffected.
	\item \textbf{Data-driven uncertainty quantification}: This research utilizes a recorded traffic dataset to characterize nominal candidate maneuvers of SVs in the IAIMM-KF, and a novel online sampling-based method to quantify trajectory uncertainty of the predicted maneuvers by the IAIMM-KF. The trajectory-uncertainty quantification approach is non-conservative and effectively captures the uncertainty properties of the vehicles’ trajectories from real-world traffic datasets.
	\item \textbf{Efficient evaluations in multi-vehicle environments:} The method demonstrates its performance in both multi-vehicle simulation scenarios and a scenario from the recorded traffic dataset with comparisons to a scenario
	MPC (SCMPC), a deterministic MPC, and a trajectory of a human-driven vehicle. Through simulations, it is shown that the method is real-time implementable.
\end{enumerate}

\subsection{Related Research} \label{Ralated Work}

\subsubsection{Motion Prediction}\label{related work prediction}
Motion prediction is important for motion planning in dynamic traffic environments, as the motion planner relies on the predicted trajectories of SVs to make decisions and to construct safety constraints. Motion-prediction models can be categorized into three levels of abstraction and complexity: physics-based models, maneuver-based models, and interaction-aware models \cite{lefevre2014survey}. Physics-based models use physical laws to predict the motion of SVs \cite{dixit2019trajectory}, \cite{fors2022resilient}. Maneuver-based models enhance the prediction by considering the possible maneuvers that the predicted SV may perform while accounting for factors like road structures \cite{carvalho2014stochastic}, \cite{nair2022stochastic}. On the other hand, interaction-aware motion models represent a more advanced approach that captures the mutual dependencies between the predicted SV and its surrounding traffic \cite{lefkopoulos2021interaction}, \cite{schulz2018interaction}, \cite{liu2021interaction}. Each type of model can be implemented through various techniques, which are further classified and summarized in \cite{lefevre2014survey}, \cite{huang2022survey}.

\subsubsection{Motion Planning}\label{related work planning}
Model predictive control (MPC) has been widely applied for motion planning of autonomous vehicles, the main reason being the inherent online replanning ability that makes it suitable in dynamic environments \cite{zhang2021optimization}, \cite{mohseni2020distributed}. The MPC-based approaches can roughly be classified either as MPC with deterministic models or MPC with stochastic models \cite[page 9-11]{rawlings2017model}. In motion-planning problems, MPC with deterministic models assumes that the modeling of the EV and the prediction of SVs are accurate such that a deterministic optimal control problem (OCP) can be solved to generate the reference trajectory. For example, in \cite{nilsson2015receding} a convex MPC was proposed for motion planning of the EV subject to linear safety constraints. In \cite{liu2017path}, a mixed-integer MPC was developed for combining maneuver selection and trajectory planning. In \cite[paper II]{mohseni2021decentralized}, a nonlinear MPC with weights in the cost function tuned by a Gaussian function was developed for adaptive collision avoidance in different scenarios. The OCP in deterministic MPC is easier to formulate, while such a method may not have the desired robustness against uncertainties.

Model predictive control with uncertain models considers the modeling uncertainty of the EV or motion-prediction uncertainties of SVs that could be uni- or multi-modal. The corresponding OCP has to be subjected to tight constraints based on the quantification of uncertainties \cite{brudigam2022safety}. Some established MPCs, like stochastic MPC (SMPC) and robust MPC (RMPC), can technically handle the motion-planning problem concerning modeling uncertainty of the EV \cite{carvalho2015automated}. For example, \cite{carvalho2014stochastic} proposed an SMPC to account for the linearization error of the vehicle model, and \cite{dixit2019trajectory} modeled the EV by a linear time-invariant model with additive uncertainty such that a convex RMPC was achieved for autonomous overtaking. 

The problem with uni-modal motion uncertainty of SVs essentially has similar complexity as the problems studied in \cite{carvalho2014stochastic} and \cite{dixit2019trajectory}. For example, \cite{fors2022resilient} used a branching MPC to handle the longitudinal acceleration uncertainty of SVs using open-loop adversarial disturbance sequences. Reference \cite{lefkopoulos2021trajectory} applied a mixed-integer SMPC to handle an obstacle with an uncertain position and unknown distribution, and a data-driven approach to estimate the distribution to formulate a tight chance constraint. In \cite{gao2022risk}, a risk-aware RMPC was formulated for an autonomous EV to safely overtake a human-driven leading vehicle that has uncertain longitudinal accelerations. 

Considering multi-modal motion uncertainties of SVs in MPC is more challenging because both the maneuver and trajectory uncertainties have to be considered. Since one mode in a multi-modal model can be described as one scenario, the scenario approaches, including both the sequence optimization-based approach \cite{cesari2017scenario, ahn2021safe, brudigam2018combining,benciolini2023non} and the policy optimization-based approach \cite{chen2022interactive,wang2022interaction,liu2021interaction,batkovic2021robust,nair2022stochastic,nair2022stochasticdual}, are naturally applied to handle the problem. The sequence-based approach computes an optimal sequence of control input for the EV over the horizon with the obstacle occupancy constructed by a scenario approach, while the policy-based approach optimizes multiple policies associated with each of the multi-modal obstacle uncertainties. For example, SCMPC was used both in \cite{cesari2017scenario} and \cite{ahn2021safe} to formulate the safety constraint considering SVs' multi-modal uncertainties; using a sampling approach the former formulated each modal by a convex hull, and the latter approximated it using a polytope. Reference \cite{brudigam2018combining} used the scenario approach to construct ellipses to cover possible maneuvers of the SV. Reference \cite{benciolini2023non} combined the output of a multi-modal motion-prediction model of SVs and an SMPC for non-conservative motion planning of an autonomous EV by leveraging the predicted maneuvers' probabilities to tune the safety level of the chance constraints. In \cite{chen2022interactive} a risk-aware branch MPC for motion planning of the ego system was proposed where the maneuver-level multi-modal uncertainties of an obstacle, which were predicted by considering the collision-avoidance with the ego system, were used to determine the weight of the branches. A related approach was presented in \cite{wang2022interaction} for considering a single SV using a learning-based branch MPC, which jointly estimated the distribution of the SV's maneuvers (tracking or braking) and optimized the control policy of the EV. An interaction-aware scenario MPC was proposed in \cite{liu2021interaction} for an autonomous EV in forced merging scenarios, where multi-modal longitudinal intentions (yielding or proceeding) of multiple SVs in the presence of the merging EV were considered in the stochastic OCP of the MPC. Reference \cite{batkovic2021robust} optimized robust policies of an SCMPC by assuming a robust minimum positive invariant set was available for each modal of the SV. Reference \cite{nair2022stochastic} optimized multiple policies associated with multi-modal uncertainties of an SV by solving a convex stochastic SCMPC, and \cite{nair2022stochasticdual} extended the approach with consideration of interactions with an SV. 

\subsection{Organization and Notation}
\paragraph*{Organization} Section~\ref{Problem Formulation} formulates the general motion-planning problem as three sub-problems, where the methods for solving each sub-problem are introduced in Section~\ref{Overall Methods}. Section~\ref{Motion Prediction} reviews the IAIMM-KF for interaction-aware multi-modal maneuver predictions of SVs, and then elaborates on the trajectory-uncertainty quantification approach. An MPC integrated with the predicted multi-modal uncertainties is presented in Section~\ref{Motion Planning}. In Section~\ref{Simulations and Discussions}, the method is verified in simulations, and Section~\ref{Conclusions} concludes the paper. 

\paragraph*{Notation} $\mathcal{R}^n$ is the $n$-dimensional real-number vector space, $\mathcal{R}_{+}^{n}$ is the $n$-dimensional non-negative vector space, $\mathcal{R}_{++}^{n}$ is the $n$-dimensional positive vector space, $0^{m, n}$ means the 0 matrix of size $m \times n$, $I^n$ means the identity matrix of size $n \times n$, $\rm{diag}$($\cdot$) means the diagonal matrix with specified elements along the diagonal, $\lfloor \cdot \rfloor$ means the round-down operation, and $\lceil \cdot \rceil$ means the round-up operation. The set addition (Minkowski sum) is represented by $\oplus$. Given a random variable $r$, $\mathbb{E}(r)$ means its expected value, and ${\rm std}\left( \left\{r^i \right\} \right)$ means the sample standard deviation of the set $\left\{r^i \right\}$, where $i$ indicates the sampling index. The parameter $N$ means the prediction horizon (both in IAIMM-KF and MPC), the current time step is indicated as $k$, and the time steps over the prediction horizon are indicated by $t$, where $t = k+1, \cdots, k + N$. The sampling interval (both in IAIMM-KF and MPC) is represented by $T$.

\section{Problem Formulation and Assumptions}\label{Problem Formulation}
\begin{figure}[t]
	\centering
	\includegraphics[width = \columnwidth]{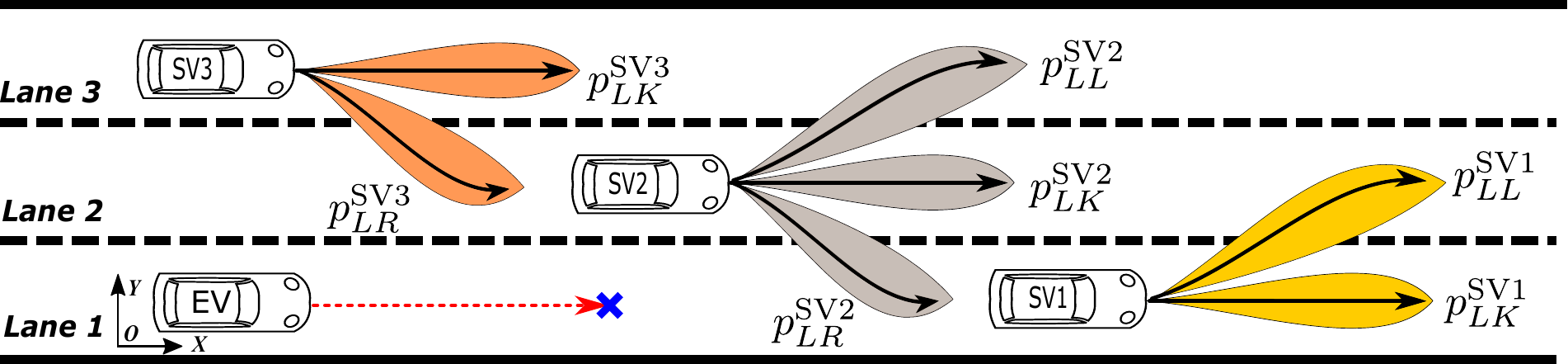}
	\caption{Scenario description, for SVs solid lines mean the predicted nominal trajectories, colored areas indicate the trajectory uncertainties, the blue cross means the reference target of the EV at the current time step, and the red dashed line is the reference trajectory of the EV by following the target.}
	\label{Fig_scenario_illustration}
\end{figure}

The problem is formulated as planning an optimal trajectory for the EV at every time step $k$ based on predicting the interactions between SVs with multi-modal uncertainties. Fig.~\ref{Fig_scenario_illustration} presents the problem in a scenario where vehicles implicitly interact with each other, and each SV has multi-modal motion uncertainties. In Fig.~\ref{Fig_scenario_illustration}, $p_{LK}^{\rm SV1}, \ p_{LL}^{\rm SV1}, \ \ldots, \ p_{LK}^{\rm SV3}$ indicate the probabilities of SVs' maneuvers, e.g., $p_{LL}^{\rm SV2}$, $p_{LK}^{\rm SV2}$, and $p_{LR}^{\rm SV2}$ mean the probability of lane-changing to the left, lane-keeping, and lane-changing to the right maneuvers of SV2. 

The motion-planning mission of the EV in the scenario of Fig.~\ref{Fig_scenario_illustration} can be divided into three tractable sub-problems: (i) Predicting the interactive maneuvers of SVs and estimating the maneuver uncertainties as illustrated by Fig.~\ref{Fig_scenario_illustration}, where the maneuver uncertainty of SVs is characterized by the nominal trajectory and the probability of each maneuver; (ii) Quantifying the trajectory uncertainty of each maneuver over the prediction horizon, i.e., quantifying the colored areas in Fig.~\ref{Fig_scenario_illustration}; (iii) Integrating the maneuver and trajectory uncertainties to compute the reference target and reference trajectory for the EV. The initial two sub-problems pertain to the motion-prediction level, while the final problem pertains to the motion-planning level. The overall methods for solving these problems are introduced in Section~\ref{Overall Methods}.

The general assumptions for this paper are: 
\begin{assumption}\label{assum: accurate observation}
	The EV has accurate observations of the SVs at the current time step $k$
\end{assumption}
\begin{assumption}\label{assum: lane change target}
	When an SV is changing lanes, the target lateral position is close to the center line of the target lane.
\end{assumption}

\section{Methods}\label{Overall Methods}
\begin{figure}[t]
	\centering
	\includegraphics[width = 0.9\columnwidth]{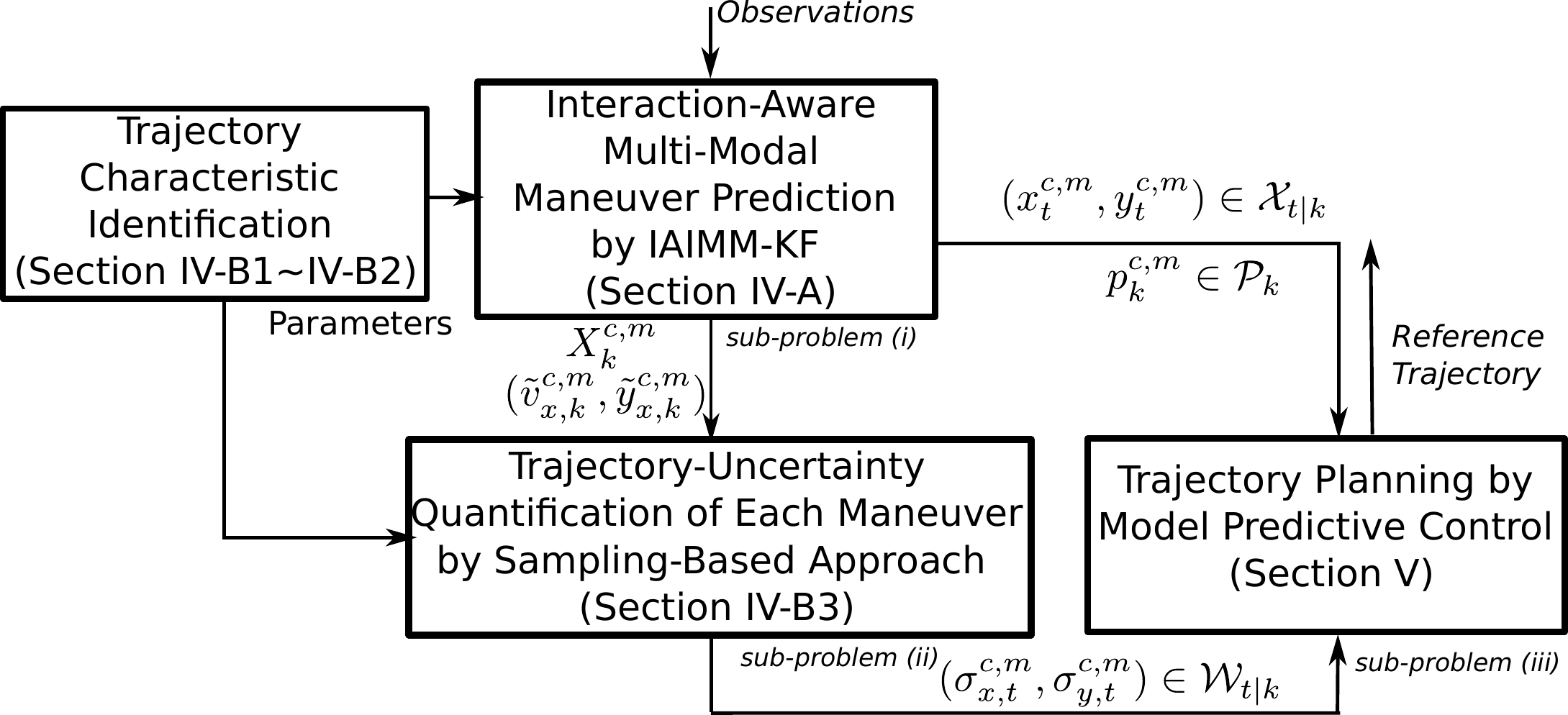}
	\caption{System architecture.}
	\label{Fig_System_Architecture}
\end{figure}

The system architecture is shown in Fig.~\ref{Fig_System_Architecture}. Based on the observations of the SVs at time step $k$, the IAIMM-KF predicts their interactive behaviors and outputs the probability $p_k^{c,m}$ and nominal longitudinal and lateral positions $(x_t^{c, m}, y_t^{c, m})$ of each possible maneuver of every SV, where $m$ is the index of the maneuver and $c$ is the index of the SV. The results $(x_t^{c, m}, y_t^{c, m})$  and $p_k^{c,m}$ are added to the sets $\mathcal{X}_{t|k}$ and $\mathcal{P}_k$, respectively. Then, for each maneuver of every SV, a sampling-based approach is applied to estimate the standard deviation (STD) of the longitudinal and lateral positions based on the measured state $X_k^{c, m}$, the reference velocity $\tilde{v}_{x,k}^{c, m}$, and reference lane position $\tilde{y}_k^{c, m}$ for maneuver $m$ of the SV $c$, where $\tilde{v}_{x,k}^{c, m}$ and $\tilde{y}_k^{c, m}$ are inferred by the IAIMM-KF. The estimated longitudinal and lateral STDs are denoted as $\sigma_{x, t}^{c, m}$ and $\sigma_{y, t}^{c, m}$, respectively, which are added to the set $\mathcal{W}_{t|k}$. The parameters that reflect the characteristics of executing maneuver $m$ of a vehicle are identified offline using a recorded traffic dataset (the highD dataset \cite{krajewski2018highd}), and they are used for the IAIMM-KF and uncertainty quantification. The parameter-identification method is elaborated on in Section~\ref{Identifying Gains Set}.

The information in the sets $\mathcal{X}_{t|k}$, $\mathcal{P}_k$, and $\mathcal{W}_{t|k}$ are integrated to formulate the OCP in an MPC for the optimal motion planning at time step $k$. Since these sets contain information concerning the interactions and uncertainties of SVs, the MPC formulated based on the sets is interaction and safety-aware. The general OCP formulation is:  
\begin{subequations}
	\begin{align}
		\mathop{\rm minimize}\limits_{U_{t-1}, \ \rho} \quad & V\left(\xi_{t|k}, U_{t-1}, \ \rho, \ \tilde{v}^{\rm EV}_{x, k}, \ \tilde{y}^{\rm EV}_k \right) & \label{eq_ocp_a} \\
		{\rm subject\ to}\quad
		&\xi_{t|k} = f(\xi_{t-1|k}, \ U_{t-1}) & \label{eq_ocp_b} \\
		&\xi_{t|k} \in \Xi  &\label{eq_ocp_c}\\
		&U_{t-1} \in \mathcal{U} &\label{eq_ocp_d}\\
		&g\left(\xi_{t|k}, \ \mathcal{X}_{t|k}, \ \mathcal{W}_{t|k}, \ \mathcal{P}_k, \ \varepsilon_k, \ \rho\right) \leq 0 & \label{eq_ocp_e} \\
		&\rho \in \mathcal{R}_{+}^{N}, \ t = k+1, \ldots, k+N & \label{eq_ocp_f}
	\end{align}
\end{subequations}
\noindent where in the cost function \eqref{eq_ocp_a}, $\xi_{t|k}$ is the EV's state predicted by a vehicle model (constraint~\eqref{eq_ocp_b}), $U_{t-1}$ is the control-input vector of the model, $\rho$ is a slack variable vector, $\tilde{v}^{\rm EV}_{x, k}$ and $\tilde{y}^{\rm EV}_k$ are the target speed and target lane position for the EV at time step $k$, respectively. The admissible sets for the vehicle state and control input are represented as $\Xi$ and $\mathcal{U}$, respectively. The safety constraint \eqref{eq_ocp_e} is formulated using the information in $\mathcal{X}_{t|k}$, $\mathcal{W}_{t|k}$, and $\mathcal{P}_k$. A pre-defined safety-awareness parameter $\varepsilon_k \in (0, 1]$ is used for tuning a trade-off between the performance and robustness of the planner. Slack variables $\rho$ enhance the feasibility of the OCP in situations where a constant collision-avoidance limit is not feasible, e.g., it handles the case if an SV cuts in directly in front of the EV.

Formulating and solving the OCP is the key part of this research, where the major focus is the targets $\tilde{v}^{\rm EV}_{x, k}$ and $\tilde{y}^{\rm EV}_k$ and the safety constraint \eqref{eq_ocp_e}, which enforce the interactive behaviors and multi-modal uncertainties of the SVs. The targets are called moving targets as they are updated at every time step to obtain non-local trajectory replanning (Section~\ref{Moving Target}). Constraint \eqref{eq_ocp_e} means the EV and SVs should keep a safe distance, and the explicit form of \eqref{eq_ocp_e} is given in Section~\ref{MPC OCP}. The OCP formulation is designed to achieve resilient motion planning by updating the references and safety constraints to accommodate dynamic traffic uncertainties and unexpected vehicle behavior. The simulations presented in Section~\ref{Simulations and Discussions} demonstrate the method's ability to adapt to unexpected uncertainties effectively. 

\begin{remark}
	Note that the maneuver-uncertainty prediction method depicted in Fig.~\ref{Fig_System_Architecture} is not limited to IAIMM-KF. Other interaction-aware prediction methods, such as learning-based approaches \cite{schulz2018interaction}, \cite{westny2023mtp}, might be employed, provided that they can predict those uncertainty parameters as input to the MPC planner. The IAIMM-KF in Fig.~\ref{Fig_System_Architecture} is preferred because of the computational efficiency. 
\end{remark}

\section{Interaction-Aware Multi-Modal Motion Prediction of Surrounding Vehicles}\label{Motion Prediction}
This section focuses on the methods for interaction-aware multi-modal motion prediction of SVs, including IAIMM-KF for predicting the maneuver uncertainties considering the interactions between SVs (Section~\ref{Prediction of Maneuver Uncertainties by IAIMM-KF}) and the sampling-based approach to quantify the trajectory uncertainties of each candidate maneuver of every SV, given the measured current state and the predicted reference state of the maneuver (Section~\ref{Estimating Trajectory Uncertainty}). The motion prediction outputs $\mathcal{X}_{t|k}$, $\mathcal{W}_{t|k}$, and $\mathcal{P}_k$, as introduced in \eqref{eq_ocp_e}, which will be utilized as input to the MPC for motion planning in Section~\ref{Motion Planning}.

\subsection{Prediction of Maneuver Uncertainties by IAIMM-KF}\label{Prediction of Maneuver Uncertainties by IAIMM-KF}

The IAIMM-KF, which was proposed as a model-based interaction-aware motion-prediction method for autonomous vehicles in \cite{lefkopoulos2021interaction}, has been used in \cite{zhou2022interaction} for modeling the traffic environments for a deterministic MPC motion planner. This section addresses three key components of the IAIMM-KF, namely, the priority list, the sub-models of SVs, and the prediction output. Theoretical details and the complete algorithm of the IAIMM-KF are given in \cite{lefkopoulos2021interaction}, and the implementation of the algorithm is available in our published code\footnote{\scriptsize\url{https://github.com/JianZhou1212/interaction-safety-aware-motion-planning}}.

\subsubsection{Priority List}\label{Priority List}
A dynamic priority list ranks the vehicles in IAIMM-KF by a descent priority such that the motion prediction is performed sequentially by the priority. The list contains all higher prioritized cars than the EV. The car with a lower priority has to consider collision-avoidance interactions with higher prioritized cars. The list at time step $k$ is denoted as $\mathcal{C}_k$, which is sorted by the following strategy \cite{lefkopoulos2021interaction}:
\begin{itemize}
	\item The leading car has a higher priority than the lagging car if two cars are in the same lane.
	\item The car with maximum longitudinal terminal position predicted with constant velocity over the prediction horizon has a higher priority if two cars are in adjacent lanes.  
\end{itemize}

According to the second rule, a longitudinally lagging vehicle may have a higher priority than a slow leading vehicle
in the adjacent lane, this is different from the case in \cite{zhou2022interaction}. Note that the construction of the priority list assumes that the EV can sense every SV within the detectable area using the onboard sensors. However, this can be challenging under a limited field of view of the vehicle. This challenge can be addressed by, e.g., integrating vehicle-to-everything (V2X) techniques.

\subsubsection{Nominal Motion Models for Surrounding Vehicles}\label{Motion Models for Surrounding Vehicles}
Deterministic models were used in \cite{zhou2022interaction} without considering uncertainties. Here, to capture maneuver uncertainties of SVs, seven different motion-primitive models are designed in this paper to describe the possible nominal maneuvers of the SVs, and the trajectory uncertainties of each maneuver are estimated separately based on the output of the IAIMM-KF. For the SVs in different lanes, the lane-dependent nominal motion primitives of the SVs are shown in Fig.~\ref{Fig_HighD_illustration}, where the maneuver set of the SVs is defined as  $\mathcal{M}_{\rm SV} = \left\{m_0, \ldots, m_6\right\}$. 

Let superscript $c$ indicate the index of the $c$-th SV in the priority list $\mathcal{C}_k$, and $m$ indicate the index of the $m$-th sub-model of the model set $\mathcal{M}_{\rm SV}$. Let $X_t^{c, m}$ denote the state vector of the SV, which contains the vehicle's longitudinal position ($x_t^{c, m}$), speed ($v_{x,t}^{c,m}$), acceleration ($a_{x,t}^{c, m}$), and lateral position ($y_t^{c, m}$), speed ($v_{y,t}^{c, m}$), acceleration ($a_{y,t}^{c, m}$) in the ground coordinate system. Assume that longitudinal and lateral accelerations are piece-wise constant over the sampling interval $T$. Then the nominal motion-primitive model is:
\begin{subequations}
	\begin{align}
		X_{t}^{c, m} &= AX_{t-1}^{c, m} + BU_{t-1}^{c, m}\label{eq_ls_a} \\
		X_t^{c, m} &= \left[
		x_{t}^{c, m} \
		v_{x, t}^{c, m} \ 
		a_{x, t}^{c, m} \
		y_{t}^{c, m} \
		v_{y, t}^{c, m} \ 
		a_{y, t}^{c, m}
		\right]^{\rm T} \label{eq_ls_b} \\
		U_t^{c, m} & = \left[
		\begin{matrix}
			-\hat{K}_{\rm lon}^m\left([v_{x,t}^{c, m}-\tilde{v}_{x, k}^{c, m} \ a_{x, t}^{c, m}]^{\rm T} \right) \\ 
			-\hat{K}_{\rm lat}^m\left([y_t^{c, m} - \tilde{y}_k^{c, m} \ v_{y, t}^{c, m} \ 
			a_{y, t}^{c, m}]^{\rm T} \right)
		\end{matrix}
		\right]  \label{eq_ls_c}	
	\end{align}
\end{subequations}
where 
$A = \left[\begin{matrix}A_x & 0^{3, 3} \\ 0^{3, 3} & A_y \end{matrix}\right], A_x = A_y = \left[\begin{matrix} 1 & T & T^2/2 \\ 0 & 1 & T \\ 0 & 0 & 1 \end{matrix}\right]$,\\
$B = \left[\begin{matrix}B_x & 0^{3, 1} \\ 0^{3, 1} & B_y \end{matrix}\right], B_x = \left[\begin{matrix} 0 \\ T^2/2 \\ T \end{matrix}\right], B_y = \left[\begin{matrix} T^3/6 \\ T^2/2 \\ T \end{matrix}\right]$. \\
The vector $U_t^{c, m}$ contains the longitudinal and lateral control inputs, which are computed by state-feedback control laws, where $\hat{K}_{\rm lon}^m \in \mathcal{R}^2$ and $\hat{K}_{\rm lat}^m \in \mathcal{R}^3$ mean the nominal longitudinal and lateral controller gains, respectively. It is defined that $\hat{K}_{\rm lon}^m = \mathbb{E}(K_{\rm lon}^m)$ and $\hat{K}_{\rm lat}^m = \mathbb{E}(K_{\rm lat}^m)$, where $K_{\rm lon}^m$ and $K_{\rm lat}^m$ are random variables with uniform distribution over the controller-gain sets $\mathcal{K}_{\rm lon}^m$ and $\mathcal{K}_{\rm lat}^m$, respectively. Parameters $\tilde{v}_{x, k}^{c, m}$ and $\tilde{y}_k^{c, m}$ mean the target speed and target lane position for maneuver $m$ of vehicle $c$, respectively. 

\subsubsection{Output of IAIMM-KF}\label{Output of IAIMM-KF}
For each maneuver $m$ of vehicle $c$, the observed signals by the IAIMM-KF are $[x_k^{c, m} \ v_{x, k}^{c, m} \ y_{x, k}^{c, m}]^{\rm T}$, which are used to predict the reference targets $\tilde{y}_k^{c, m}$ and $\tilde{v}_{x, k}^{c, m}$ by solving an optimization problem subjected to collision-avoidance constraints with all higher prioritized SVs than the SV $c$. It follows that $x_t^{c, m}$ and $y_t^{c, m}$, i.e., the nominal longitudinal and lateral positions of the maneuver at time step $t$, are predicted by tracking the references with the controller gains $\hat{K}_{\rm lon}^m$ and $\hat{K}_{\rm lat}^m$ in \eqref{eq_ls_a}--\eqref{eq_ls_c}. The maneuver probability $p_k^{c, m}$ is computed based on the innovation residuals and the optimization cost. Let $p_k^{c, m} \in \mathcal{P}_k$ and $(x_t^{c, m}, \ y_t^{c, m}) \in \mathcal{X}_{t|k}$ for use in the MPC.

The IAIMM-KF outputs the probability and the nominal trajectory of each maneuver $m$ of the SV $c$ to quantify the uncertainties on the maneuver level. However, starting from the current state $X_k^{c, m}$, the vehicle has multiple options in terms of trajectories to reach the reference targets $\tilde{y}_k^{c, m}$ and $\tilde{v}_{x, k}^{c, m}$, and this results in the trajectory uncertainty, which has to be estimated to quantify the uncertainty at the trajectory level. The details of the quantification approach are introduced in Section~\ref{Estimating Trajectory Uncertainty}.

\begin{remark}
	The IAIMM-KF is an augmented version of the interacting multiple model Kalman filter (IMM-KF). IMM-KF is a maneuver-based motion-prediction model \cite{carvalho2015automated} that has been commonly used for predicting multi-modal uncertainties of SVs in motion-planning strategies \cite{carvalho2014stochastic}, \cite{benciolini2023non}, \cite{burger2020interaction}. Compared with IMM-KF, IAIMM-KF advances in the ability in modeling the interactions between the SVs. Experimental results in \cite{zhou2022interaction} demonstrated that a motion-planning strategy using IAIMM-KF predictions is more proactive compared to the traditional IMM-KF approach.
\end{remark}

\begin{figure}[!t]
	\centering
	\includegraphics[width = \columnwidth]{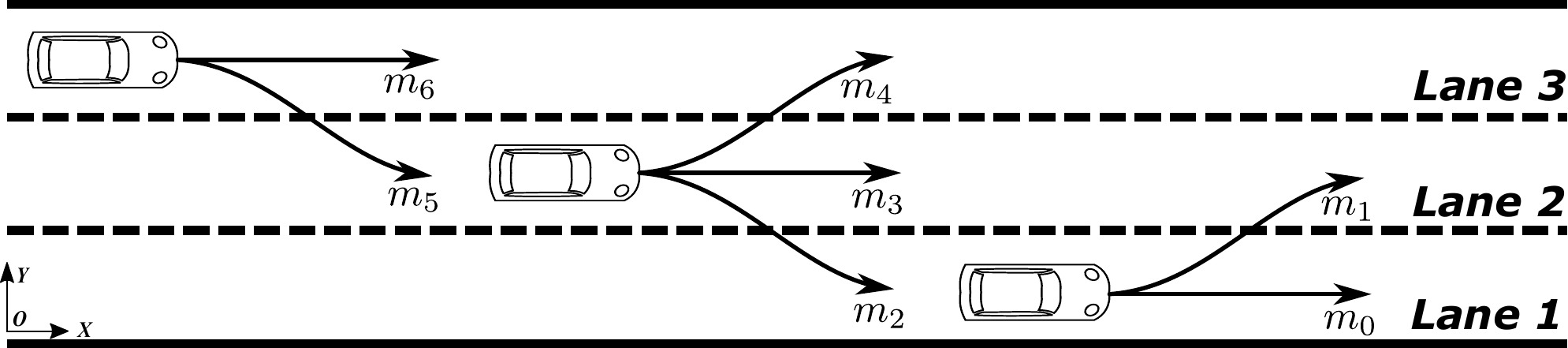}
	\caption{The nominal maneuvers of an SV in different lanes.}
	\label{Fig_HighD_illustration}
\end{figure}

\begin{remark}
\begin{itemize}
	\item Fig.~\ref{Fig_HighD_illustration} defines seven lane-dependent maneuvers for the SVs. This is different from  \cite{zhou2022interaction} and \cite{lefkopoulos2021interaction} where only three maneuvers (left lane-changing, lane-keeping, and right lane-changing) are considered. 
	\item The uncertainties of the model \eqref{eq_ls_a}--\eqref{eq_ls_c} are characterized by random controller gains $K_{\rm lon}^{m}$ and $K_{\rm lat}^{m}$, instead of an additive noise term as in \cite[Eq. (10)]{lefkopoulos2021interaction}. The benefits of using model \eqref{eq_ls_a}--\eqref{eq_ls_c} will be illustrated in Section~\ref{Online Estimating Trajectory Uncertainty}.
	\item The IAIMM-KF can be used to either simulate a traffic scenario or predict the motion of vehicles in a given traffic scenario. The simulation property will be used in Section~\ref{SIM1}, and the prediction property will be used in Section~\ref{SIM2}--\ref{SIM3}. 
\end{itemize}
\end{remark}

\begin{figure}[!t]
	\centering
	\subfloat[ ]{\includegraphics[width = \columnwidth]{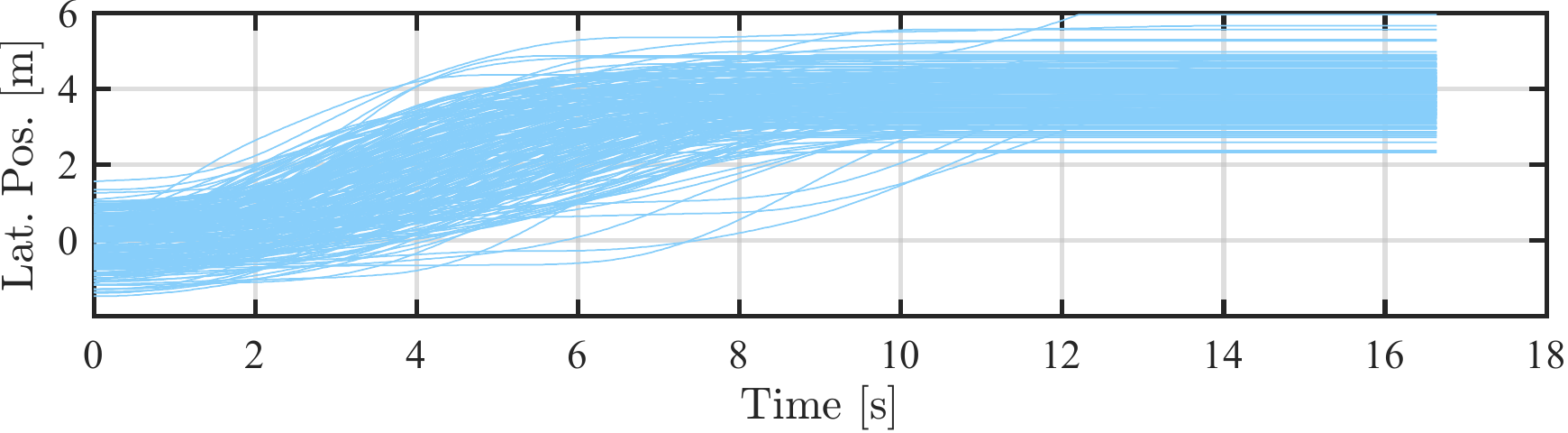}%
	\label{Fig_Normalization_y}} 
	\vspace{0 cm}
	\subfloat[ ]{\includegraphics[width = \columnwidth]{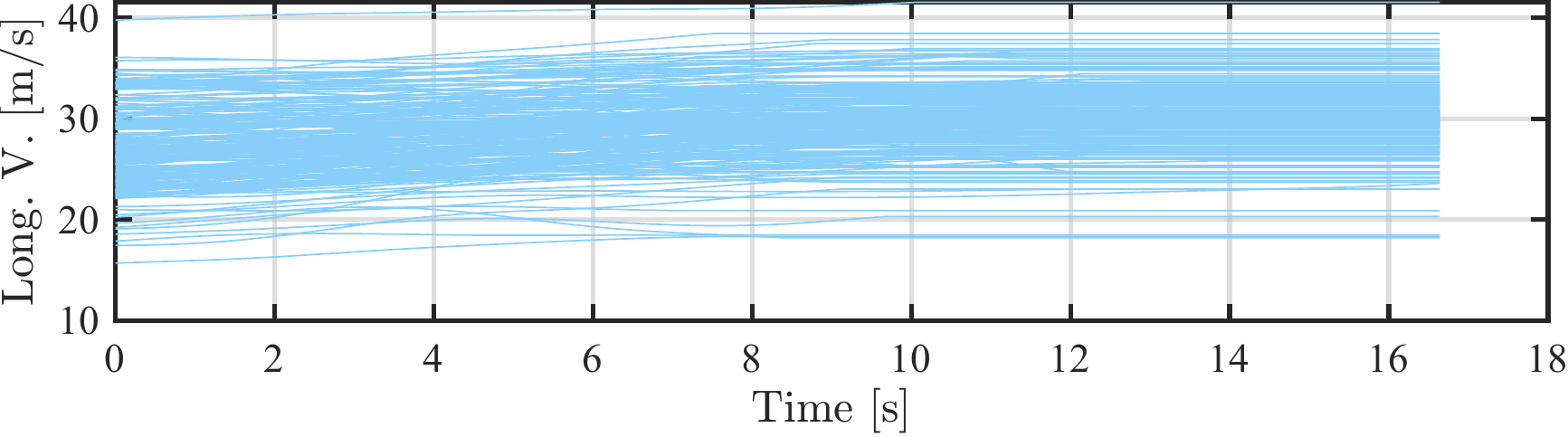}%
	\label{Fig_Normalization_v}}
	\caption{Normalized trajectory clusters of maneuver $m_1$. (a) Normalized lateral position cluster. (b) Normalized longitudinal velocity cluster.}
	\label{Fig_Normalization}
\end{figure}

\subsection{Quantification of Trajectory Uncertainty}\label{Estimating Trajectory Uncertainty}
After obtaining the nominal trajectory, the probability, and the reference targets of maneuver $m$ of vehicle $c$ as the output of the IAIMM-KF, this subsection introduces how to quantify the trajectory uncertainty of the corresponding maneuver. The quantification is performed online, based on the discrete-time controller-gain sets $\mathcal{K}_{\rm lon}^m$ and $\mathcal{K}_{\rm lat}^m$, which are identified offline using the highD dataset \cite{krajewski2018highd}. The identification is based on normalized trajectory clusters extracted from the traffic dataset recorded on three-lane roads with vehicles longitudinally driving from left to right, as indicated in Fig.~\ref{Fig_HighD_illustration}.

\subsubsection{The Normalized Trajectory Cluster}\label{Normalized Trajectory Cluster}
For each maneuver $m$, defined in Fig.~\ref{Fig_HighD_illustration}, the trajectory clusters of lateral positions and longitudinal velocities are formed using the highD dataset. The trajectories in the cluster are normalized to have the same temporal duration. This is achieved by taking the longest trajectory of the cluster as the reference. A trajectory shorter than the reference is extended by propagating from the terminal point with a constant lateral position and longitudinal velocity such that every trajectory of the cluster has the same time duration. Take maneuver $m_1$ in Fig.~\ref{Fig_HighD_illustration} for instance, the normalized trajectory cluster is shown in  Fig.~\ref{Fig_Normalization}. The trajectory clusters of the other maneuvers of the set $\mathcal{M}_{\rm SV}$ are qualitatively similar.

\subsubsection{Offline Identification of $\mathcal{K}_{\rm lon}^m$ and $\mathcal{K}_{\rm lat}^m$}\label{Identifying Gains Set}
The elements in the sets $\mathcal{K}_{\rm lon}^m$ and $\mathcal{K}_{\rm lat}^m$ characterize the variations in longitudinal and lateral driving behaviors within the same maneuver. The elements in the sets can be identified based on the normalized trajectory clusters in Fig.~\ref{Fig_Normalization}. The subscript $n$ indicates the index of the $n$-th trajectory of the cluster, and let $K_{{\rm lon}, n}^m \in \mathcal{K}_{\rm lon}^m$ and $K_{{\rm lat}, n}^m \in \mathcal{K}_{\rm lat}^m$ denote the corresponding longitudinal and lateral controller gains, respectively. The following nonlinear least squares (NLS) problem is designed to identify $K_{{\rm lat},n}^m$ based on trajectory $n$:
\begin{subequations}
	\begin{align}
		\mathop{\rm minimize}\limits_{Q_n} \quad & \sum_{j=0}^{N_m-1}||y_{n, j}^m -  H\cdot X_{{\rm lat}, n, j}^m||_2^2 & \label{eq_nls_single_a} \\
		{\rm subject\ to}\quad
		&X_{{\rm lat}, n, j+1}^m = A_yX_{{\rm lat}, n, j}^m + B_yU_{{\rm lat},n, j}^m & \label{eq_nls_single_b} \\
		&U_{{\rm lat}, n, j}^m = K_{{\rm lat}, n}^m\left(\tilde{X}_{{\rm lat}, n}^m-X_{{\rm lat}, n, j}^m\right) &\label{eq_nls_single_c}\\
		&K_{{\rm lat}, n}^m = {\rm LQR}(A_y, B_y, Q_n, R_n)&\label{eq_nls_single_d}\\
		&Q_n \in \rm{diag}(\mathcal{R}_{++}^3),  \ {\it R_n} = 1 & \label{eq_nls_single_e} 
	\end{align}
\end{subequations}
\noindent where $N_m = \left \lfloor  \tau_m/T \right \rfloor $ and $\tau_m$ is the normalized duration of the trajectory cluster. The parameter $y_{n, j}^m$ is the lateral position at time step $j$ of trajectory $n$. The vector $H = [1 \ 0 \ 0]$ extracts the lateral position from the state vector $X_{{\rm lat}, n, j}^m$, where $X_{{\rm lat}, n, j}^m$ contains the lateral position, velocity, and acceleration at time step $j$. The lateral model is propagated in \eqref{eq_nls_single_b} with $X_{{\rm lat},n,0}^m$ as the initial state of trajectory $n$, where the system matrices $A_y$ and $B_y$ are defined in \eqref{eq_ls_b}. In \eqref{eq_nls_single_c}, $\tilde{X}_{{\rm lat},n}^m$ is the reference that takes the terminal state of the trajectory $n$. The constraints \eqref{eq_nls_single_d} determines ${K}_{{\rm lat},n}^{m}$ via a linear-quadratic regulator (LQR) \cite[page 24-26]{rawlings2017model} with the diagonal of the $Q_n$ matrix as optimization variables. Estimating $K_{{\rm lat}, n}^m$ using the LQR formulation ensures a stabilizing feedback gain. 

Identifying the longitudinal controller gains $K_{{\rm lon},n}^m$ follows the same approach as in \eqref{eq_nls_single_a}--\eqref{eq_nls_single_e}. The NLS approach is performed for every trajectory in the trajectory cluster of maneuver $m$ to get the sets $\mathcal{K}_{\rm lon}^m$ and $\mathcal{K}_{\rm lat}^m$, which are used online to estimate the trajectory uncertainty of the maneuver.

\subsubsection{Online Quantification of Trajectory Uncertainty}\label{Online Estimating Trajectory Uncertainty}
\begin{figure}[!t]
	\centering
	\subfloat[ ]{\includegraphics[width = \columnwidth]{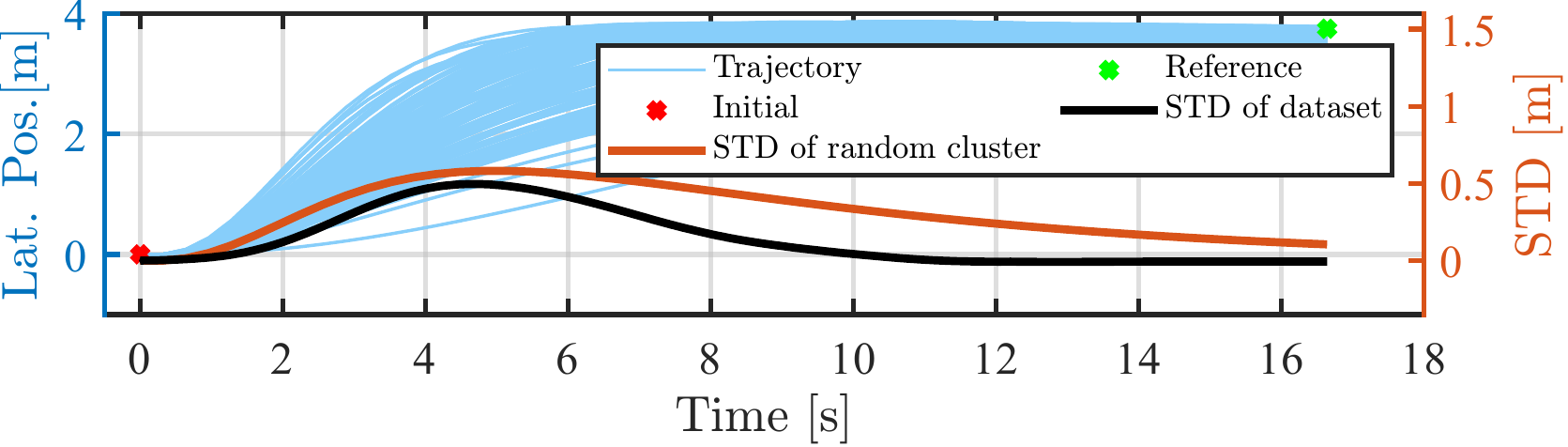}%
		\label{Fig_Cluster_Sampling_y}} 
	\vspace{0 cm}
	\subfloat[ ]{\includegraphics[width = \columnwidth]{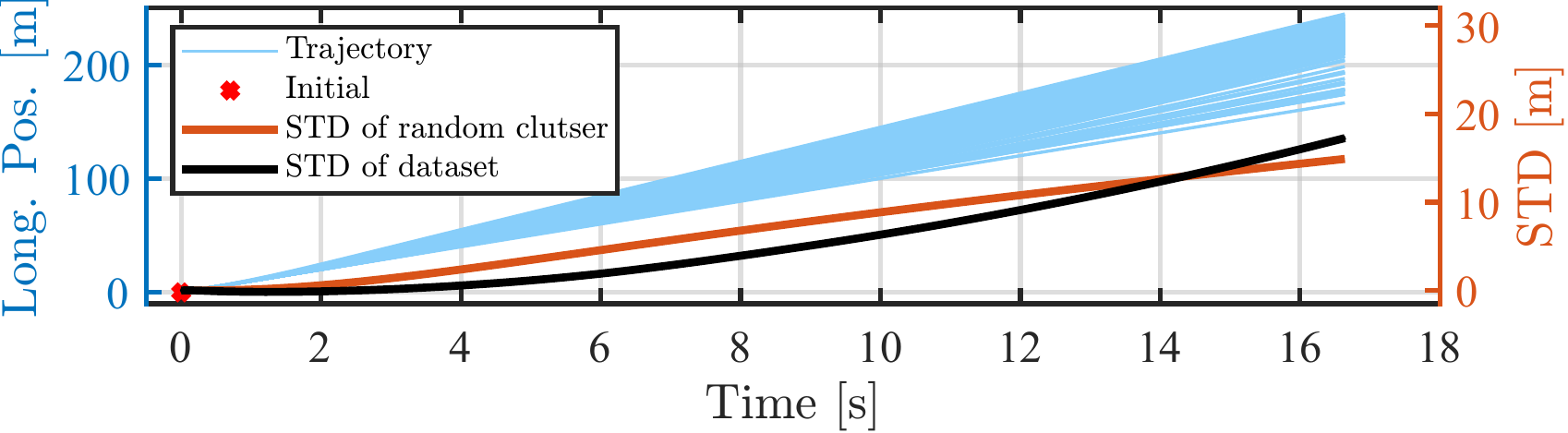}%
		\label{Fig_Cluster_Sampling_x}}
	\caption{Random trajectory clusters and STDs generated by the sampling-based approach, and the STDs of the trajectory clusters of the highD dataset. (a) The lateral position cluster and the STDs. (b) The longitudinal position cluster and the STDs. (Note that the longitudinal position cluster is generated by tracking a reference velocity).}
	\label{Fig_Cluster_Sampling}
\end{figure}

\begin{figure}[!t]
	\centering
	\includegraphics[width = \columnwidth]{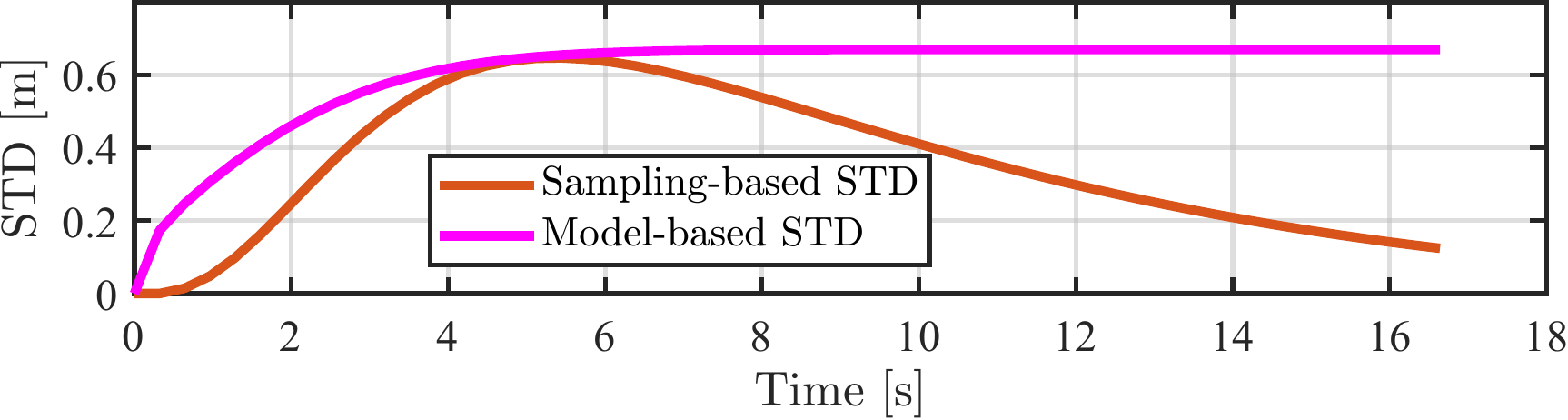}
	\caption{Comparison of the STD of the lateral lane-changing position predicted by the sampling-based approach and the model-based approach.}
	\label{Fig_STD_y_model_datadriven}
\end{figure}

The trajectory uncertainties of maneuver $m$ of SV $c$ are quantified by predicting the STDs $\sigma_{x, t}^{c, m}$ and $\sigma_{y, t}^{c, m}$ of longitudinal and lateral positions at time step $t$ over the prediction horizon. Given the sets $\mathcal{K}_{\rm lon}^m$ and $\mathcal{K}_{\rm lat}^m$, the trajectory uncertainty can be predicted by a sampling-based approach, which is similar to the scenario-generation method in \cite{schildbach2015scenario}. With the current state and the reference targets of an SV determined, according to Assumption~\ref{assum: lane change target}, the approach first samples the controller gains $K_{\rm sam}$ times from $\mathcal{K}_{\rm lon}^m$ and $\mathcal{K}_{\rm lat}^m$ and substitutes into the model \eqref{eq_ls_a}--\eqref{eq_ls_c} to generate a random trajectory cluster. It follows that $\sigma_{x, t}^{c, m} = {\rm std}\left(\left\{x_{t, i}^{c, m}\right\}\right)$ and $\sigma_{y, t}^{c, m} = {\rm std}\left(\left\{y_{t, i}^{c, m}\right\}\right)$, where $\left\{x_{t, i}^{c, m}\right\}$ and $\left\{y_{t, i}^{c, m}\right\}$ are sample sets that collect the longitudinal and lateral positions at time step $t$ of the $i$-th trajectory of the random trajectory cluster. The STDs $\sigma_{x, t}^{c, m}$ and $\sigma_{y, t}^{c, m}$ are added to the set $\mathcal{W}_{t|k}$ for describing the distribution of uncertain positions of SV $c$.

To illustrate the approach, let $K_{\rm sam}=200$, the initial velocity be $10 \ {\rm m/s}$, the reference velocity be $15 \ {\rm m/s}$, the initial longitudinal and lateral positions be ${0 \ \rm m}$, and the reference lateral position be ${3.75 \ \rm m}$. The random trajectory clusters of the lateral and longitudinal positions for maneuver $m_1$ generated by the sampling-based approach, as well as the STDs computed using the random trajectory clusters, are shown in Fig.~\ref{Fig_Cluster_Sampling}. The STD of the lane-changing trajectory cluster in Fig.~\ref{Fig_Normalization}(a), as well as the STD of the trajectory cluster of the longitudinal positions corresponding to Fig.~\ref{Fig_Normalization}(b), are also presented in Fig.~\ref{Fig_Cluster_Sampling}. The STDs of the trajectory clusters of the dataset are normalized to start from 0, based on Assumption~\ref{assum: accurate observation}. It is seen that the sampling-based approach can capture the feature of the STDs of the trajectory clusters from the highD dataset, where the STD of the lane-changing trajectories is convergent and the STD of the longitudinal trajectories is increasing over the horizon. In addition, the STDs predicted by the sampling-based approach are less conservative compared with propagating uncertainties by a model over the horizon \cite[Fig. 1 and Eq. (21)]{westny2023mtp}. This is shown in Fig.~\ref{Fig_STD_y_model_datadriven}, where the magenta line represents the STD of the lateral lane-changing position propagated based on model \eqref{eq_ls_a} with a covariance matrix that is tuned to make the two STDs in Fig.~\ref{Fig_STD_y_model_datadriven} have the same peak value. Compared with the model-based STD, the sampling-based approach reduces the conservatism of predicting the uncertainty.

For the lane-keeping maneuvers ($m \in \left\{m_0, m_3, m_6\right\}$), it is observed from the highD dataset that the STD is close to a constant, which can be estimated as the STD of the trajectory cluster of maneuver $m$. Hence, for the lane-keeping maneuvers of the SVs, a constant STD will be used over the prediction horizon to quantify the trajectory uncertainty.

Note that in the motion-planning method, these trajectory clusters are sampled online depending on the observed state and the inferred longitudinal and lateral references of the SV.

\section{Interaction and Safety-Aware MPC for Motion Planning Based on Motion Predictions}\label{Motion Planning}
This section introduces the MPC, which uses $\mathcal{X}_{t|k}$ that collects the predicted nominal trajectories of the SVs, $\mathcal{P}_k$ that collects the probabilities for all maneuvers of the SVs, and $\mathcal{W}_{t|k}$ that collects the trajectory-uncertainty parameters of the SVs, for interaction and safety-aware optimal motion planning of the EV. The MPC design enforces the general formulation \eqref{eq_ocp_a}--\eqref{eq_ocp_f}. Specifically, the EV model constraints \eqref{eq_ocp_b}--\eqref{eq_ocp_d} are specified in Section~\ref{Vehicle Model}. Section~\ref{GMM Obstacle} introduces how to use a tunable parameter $\varepsilon_k$ to design the adjustable obstacle occupancy based on motion prediction of SVs for safety awareness in motion planning. Section~\ref{Moving Target} computes the reference state for the EV in the MPC, i.e., $\tilde{v}_{x,k}^{\rm EV}$ and $\tilde{y}_k^{\rm EV}$ in \eqref{eq_ocp_a}, based on the designed obstacle occupancy. Section~\ref{Most Direct Vehicle} implements the collision-avoidance constraint \eqref{eq_ocp_e} based on the obstacle occupancy, and Section~\ref{MPC OCP} presents the specific OPC formulation for computing the reference trajectory of the EV at the current time step $k$.

\subsection{Kinematic Vehicle Model}\label{Vehicle Model}
Based on the 4:th order kinematic vehicle model used in \cite{dixit2019trajectory}, this paper uses the longitudinal snap and angular acceleration of the front wheel as the control input, which results in an 8:th order model, to get a smoother reference trajectory of the EV. The 8:th order model is \cite{zhou2022interaction}: 
\begin{subequations}
	\begin{align}
		\left(\dot x \ \dot v \ \dot a \ \dot {\eta}\ \dot \delta \ \dot \omega \right) & = \left(v \ a \ \eta \ s \ \omega \ \alpha\right) \label{eq_vehicle_model_a}\\
		\dot y & = v\phi+\frac{l_r}{l_r+l_f}v\delta \label{eq_vehicle_model_b} \\
		\dot \phi & = \frac{1}{l_f+l_r}v\delta \label{eq_vehicle_model_c} 
	\end{align}
\end{subequations}
where $x$ and $y$ are the longitudinal and lateral positions of the center of gravity in the ground coordinate system, $v$ is the longitudinal speed in the vehicle frame, $\phi$ is the yaw angle, and $l_f$, $l_r$ are the distance of the front and rear axle to the center of gravity, respectively. The front wheel angle is represented by $\delta$, the angular rate is $\omega$, and the angular acceleration is $\alpha$. The variables $a$, $\eta$, and $s$ are the longitudinal acceleration, jerk, and snap in the vehicle frame, respectively. The discrete-time form of model \eqref{eq_vehicle_model_a}--\eqref{eq_vehicle_model_c} at time step $k$ is:
\begin{eqnarray}
	\xi_{k+1}^{\rm EV} = f(\xi_{k}^{\rm EV}, U_{k}^{\rm EV}) \label{eq_vehicle_model_ode}
\end{eqnarray}
where $\xi_k^{\rm EV} = \left[ x_k^{\rm EV} \ y_k^{\rm EV} \ \phi_k^{\rm EV} \ v_k^{\rm EV} \ a_k^{\rm EV} \ \eta_k^{\rm EV} \ \delta_k^{\rm EV} \ \omega_k^{\rm EV}\right]^{\rm T}$ is the state, and $U_k^{\rm EV} = \left[s_k^{\rm EV} \ \alpha_k^{\rm EV}\right]^{\rm T}$ is the input. The relation \eqref{eq_vehicle_model_ode} is obtained by a fourth-order Runge-Kutta method with the sampling interval {\it T}, and $s_k^{\rm EV}$ and $\alpha_k^{\rm EV}$ are constant within $T$.

\subsection{The Safety-Aware Occupancy of Surrounding Vehicles}\label{GMM Obstacle}
For each maneuver of the SVs, we have predicted the maneuver probability and nominal positions over the prediction horizon and estimated the STDs of the uncertain positions. Therefore, the distribution of the uncertain position of each SV at every time step over the prediction horizon can naturally be approximated by a Gaussian mixture model (GMM). For the SV $c \in \mathcal{C}_k$, the probability density function (PDF) of the GMM at time step $t$ is defined as:
\begin{equation}
	f_t^c(x, y) = \sum_{m \in \mathcal{M}}\frac{p_k^{c, m} \cdot \mathcal{N}\left((x, y)|{\mu}_t^{c, m}, {\Sigma}_t^{c, m}\right)}{{\rm max}\left(\mathcal{N}\left((x, y)|\mu_t^{c, m}, \Sigma_t^{m}\right)\right)} \label{eq_normalize_pdf}
\end{equation}
\noindent where $x_{\rm min} \leq x \leq x_{\rm max}, \ y_{\rm min} \leq y \leq y_{\rm max}$, defining the truncated support for the PDF at time step $t$, considering that the position uncertainty cannot go to infinity in practice \cite{wang2020non}. The mean of the normal distribution $\mathcal{N}$ is represented as $\mu_t^{c, m} = [x_t^{c,m} \ y_t^{c,m}]^{\rm T}$ and the covariance matrix is $\Sigma_t^{m} = {\rm diag}\left(\left[(\sigma_{x, t}^{c, m})^2 \ (\sigma_{y, t}^{c, m})^2\right]\right)$ with $(x_t^{c,m}, y_t^{c,m}) \in \mathcal{X}_{t|k}$ and $(\sigma_{x, t}^{c, m}, \sigma_{y, t}^{c, m}) \in \mathcal{W}_{t|k}$. Note that \eqref{eq_normalize_pdf} is reshaped by the denominator such that the height of the peak in the PDF is proportional to the mode probability $p_k^{c, m}$, where $p_k^{c, m} \in \mathcal{P}_k$.

The full support of $f_t^c(x, y)$ defines the most conservative occupancy of the SV at time step $t$, where the worst case of position uncertainty of the SV is included. However, using the full support of $f_t^c(x, y)$ in the robust MPC planner would be overly conservative, such that the performance of the planned trajectory will be reduced \cite{gao2022risk}. In this paper, the tunable parameter $\varepsilon_k$ in \eqref{eq_ocp_e} is used to find a less conservative boundary of the occupancy within the support of $f_t^c(x, y)$, which is denoted as $\mathcal{D}_t^c$, to achieve a trade-off between robustness and performance of the motion-planning method. Let $f_{t, {\rm max}}^c = {\rm max}\left\{f_t^c(x, y)\right\}$, $\mathcal{D}_t^c$ is then computed as: 
\begin{equation}
	\mathcal{D}_t^c = \left\{ (x, y) \ | \ f_t^c(x, y) = \varepsilon_k \cdot f_{t, {\rm max}}^c \right\} \label{eq_intercept_pdf}
\end{equation}

Let
\begin{align*} 
	\underline{\mathcal{D}}_{t, x}^c & = {\rm min}\left\{x \ | \ \exists y, (x, y) \in \mathcal{D}_t^c\right\},\\
	\overline{\mathcal{D}}_{t, x}^c & = {\rm max}\left\{x \ | \ \exists y, (x, y) \in \mathcal{D}_t^c\right\},  \\
	\underline{\mathcal{D}}_{t, y}^c & = {\rm min}\left\{y \ | \ \exists x, (x, y) \in \mathcal{D}_t^c\right\}, \\ 
	\overline{\mathcal{D}}_{t, y}^c & = {\rm max}\left\{y \ | \ \exists x, (x, y) \in \mathcal{D}_t^c\right\}.
\end{align*} 
The minimal rectangular occupancy $\mathcal{O}_t^c$ that covers all points on and within the boundary $\mathcal{D}_t^c$ is defined as: 
\begin{equation}
		\mathcal{O}_t^c = \left\{ (x, y) \ | \ \underline{\mathcal{D}}_{t, x}^c \leq x \leq \overline{\mathcal{D}}_{t, x}^c, \
				\underline{\mathcal{D}}_{t, y}^c \leq y \leq \overline{\mathcal{D}}_{t, y}^c \right\} \label{eq_gmm_obstacle}
\end{equation}

\begin{figure}[!t]
	\centering
	\includegraphics[width = 0.7\columnwidth]{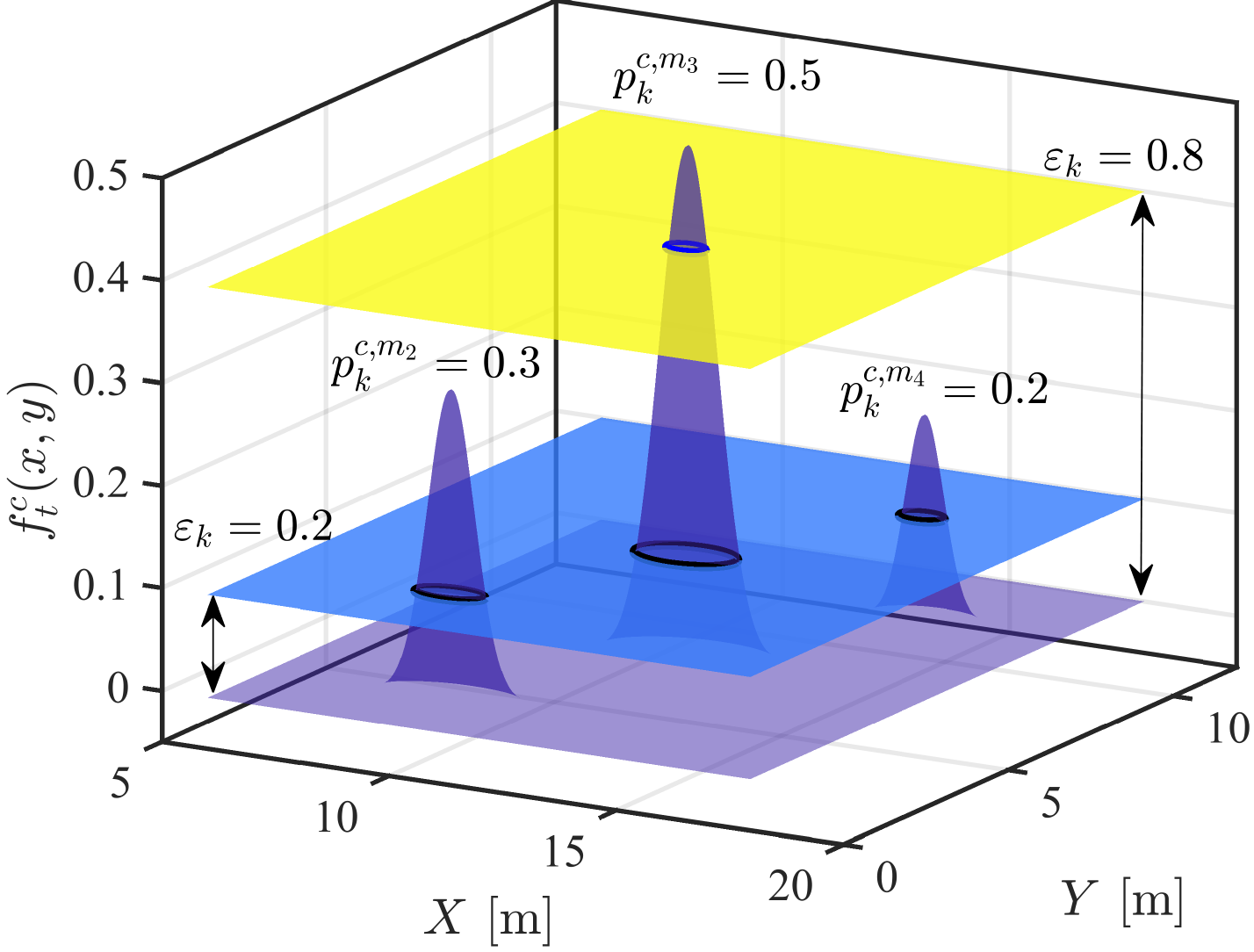}
	\caption{The intersection between $\varepsilon_k\cdot f_{t, {\rm max}}^c$ and $f_t^c(x, y)$.}
	\label{Fig_intercept_GMM}
\end{figure}

\begin{figure}[!t]
	\centering
	\subfloat[ ]{\includegraphics[width = 0.5\columnwidth]{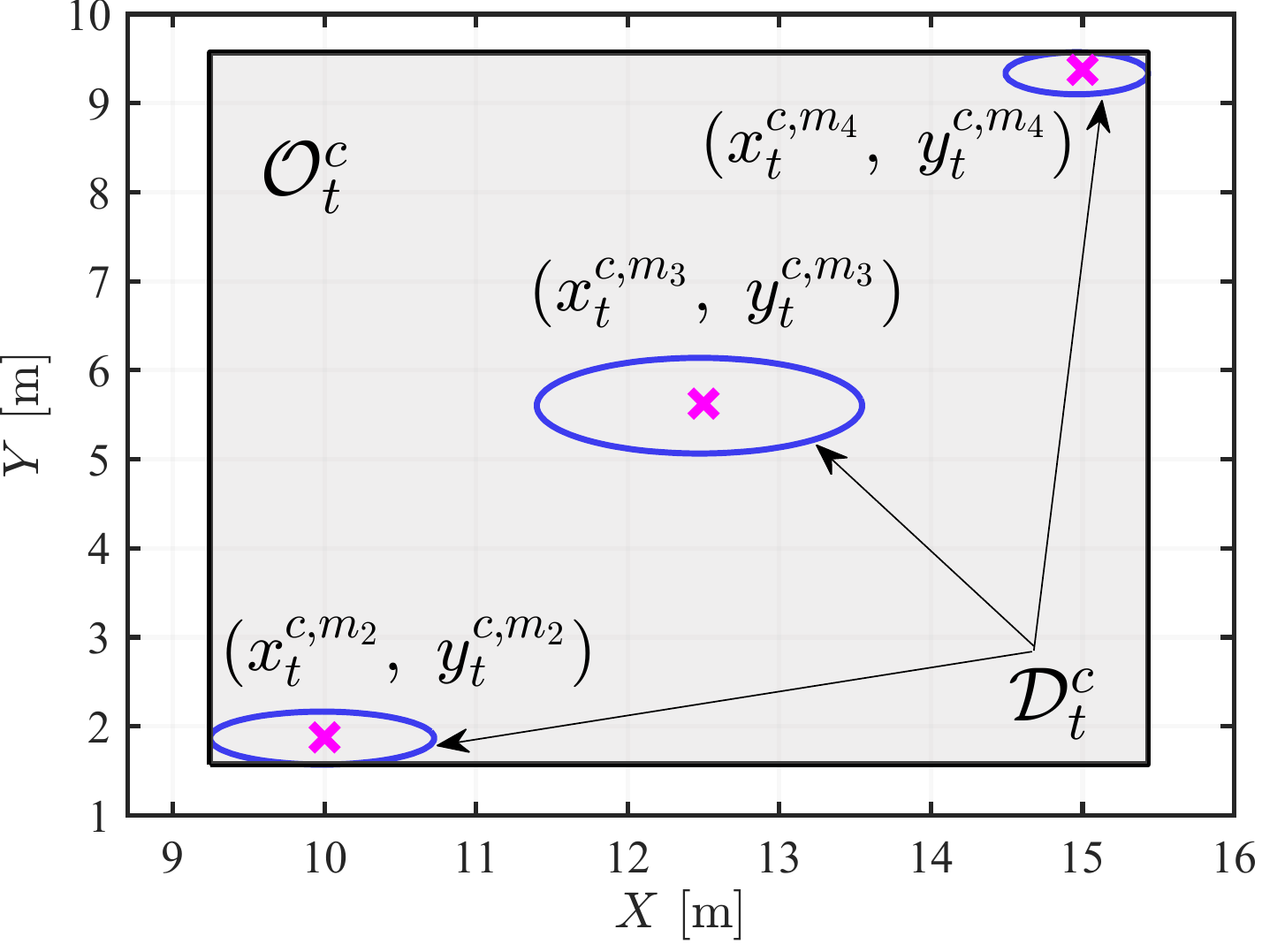}%
		\label{Fig_GMM_obstacle_a}} 
	\hfil
	\subfloat[ ]{\includegraphics[width = 0.5\columnwidth]{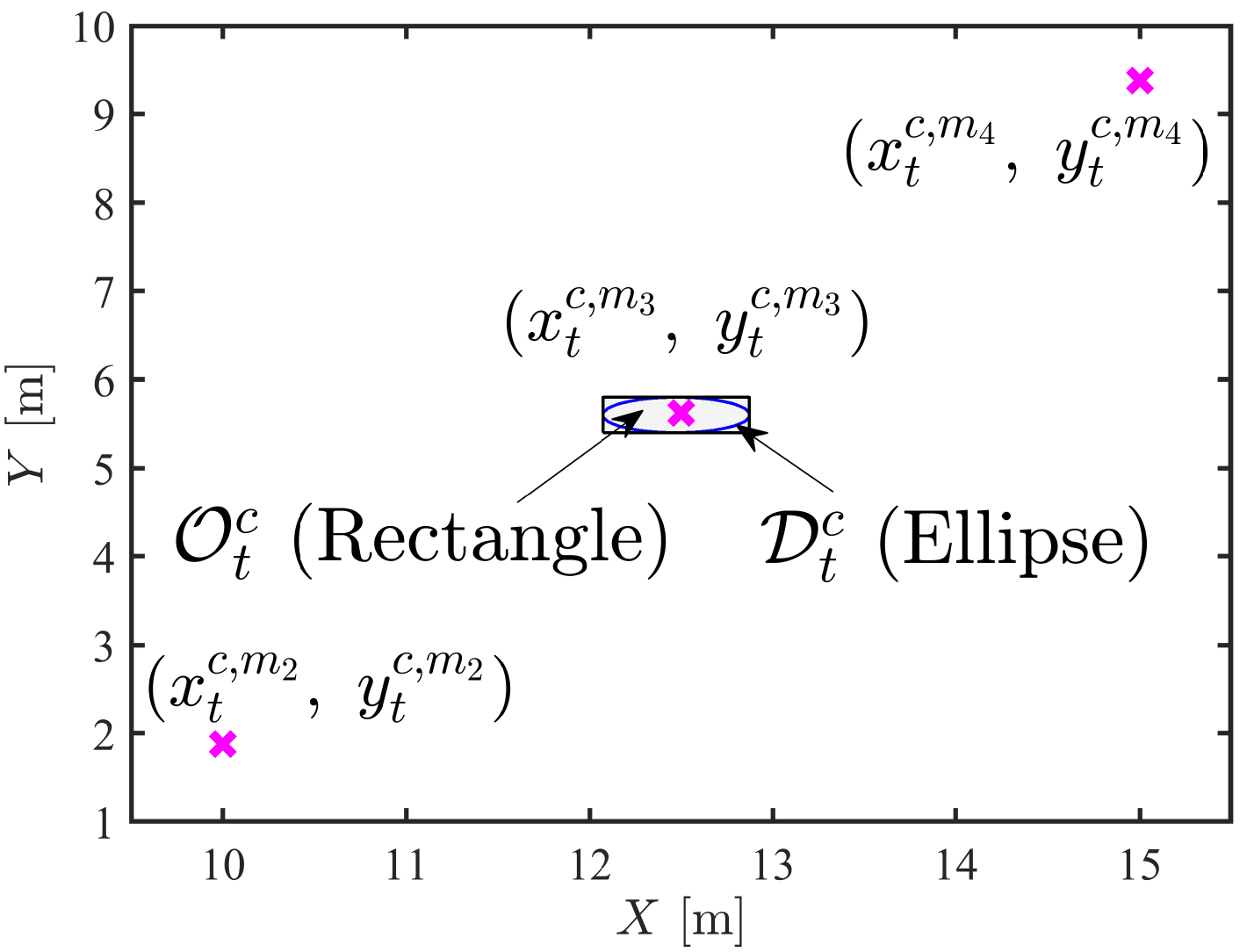}%
		\label{Fig_GMM_obstacle_b}}
	\caption{The sets $\mathcal{D}_t^c$ and $\mathcal{O}_t^c$ with different values of $\varepsilon_k$. (a) Parameter $\varepsilon_k = 0.2$. (b) Parameter $\varepsilon_k = 0.8$.}
	\label{Fig_GMM_obstacle}
\end{figure}

To visualize $\mathcal{D}_t^c$ and $\mathcal{O}_t^c$, assume that the SV $c$ is in lane 2 in Fig.~\ref{Fig_HighD_illustration} such that three maneuvers $m_2, \ m_3, \ m_4$ are available. Let the maneuver probabilities computed by IAIMM-KF be $p_k^{c, m_2} = 0.3, \ p_k^{c, m_3} = 0.5, \ p_k^{c, m_4} = 0.2$, and let $\varepsilon_k$ be 0.2 and 0.8, respectively. The ellipses in Fig.~\ref{Fig_intercept_GMM} are defined by intersecting $\varepsilon_k\cdot f_{t, {\rm max}}^c$ with $f_t^c(x, y)$ according to \eqref{eq_intercept_pdf} with different values of $\varepsilon_k$. These ellipses are projected onto the $X$-$Y$ plane to form the set $\mathcal{D}_t^c$, as shown in Fig.~\ref{Fig_GMM_obstacle}, where the gray areas indicate the set $\mathcal{O}_t^c$ defined in \eqref{eq_gmm_obstacle}.
\begin{figure}[!t]
	\centering
	\subfloat[ ]{\includegraphics[width = \columnwidth]{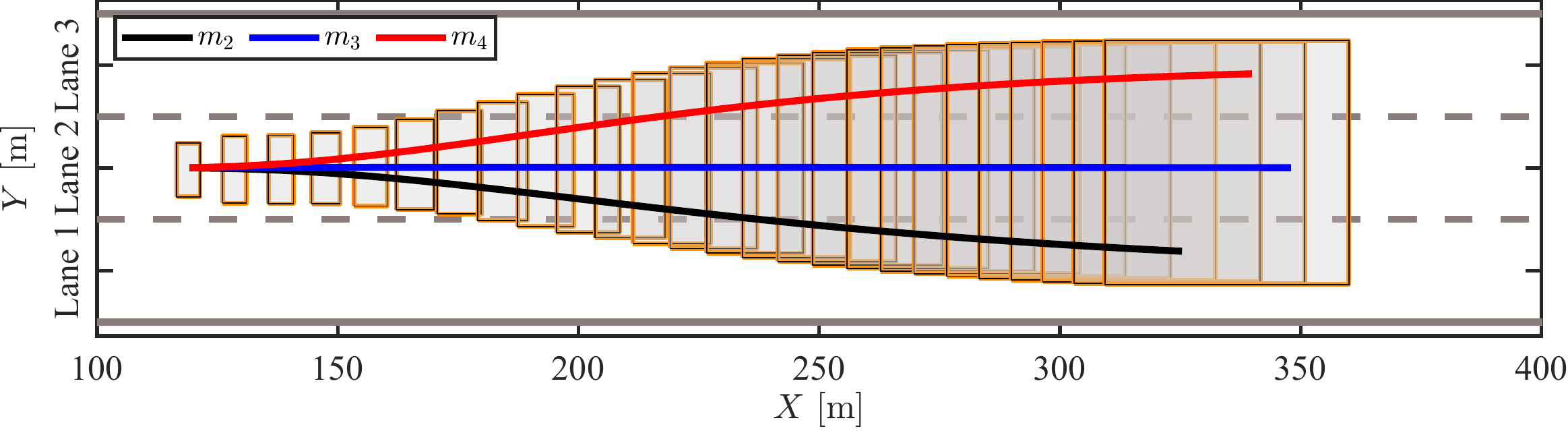}%
		\label{Fig_GMM_obst_road__a}} 
	\vspace{0 cm}
	\subfloat[ ]{\includegraphics[width = \columnwidth]{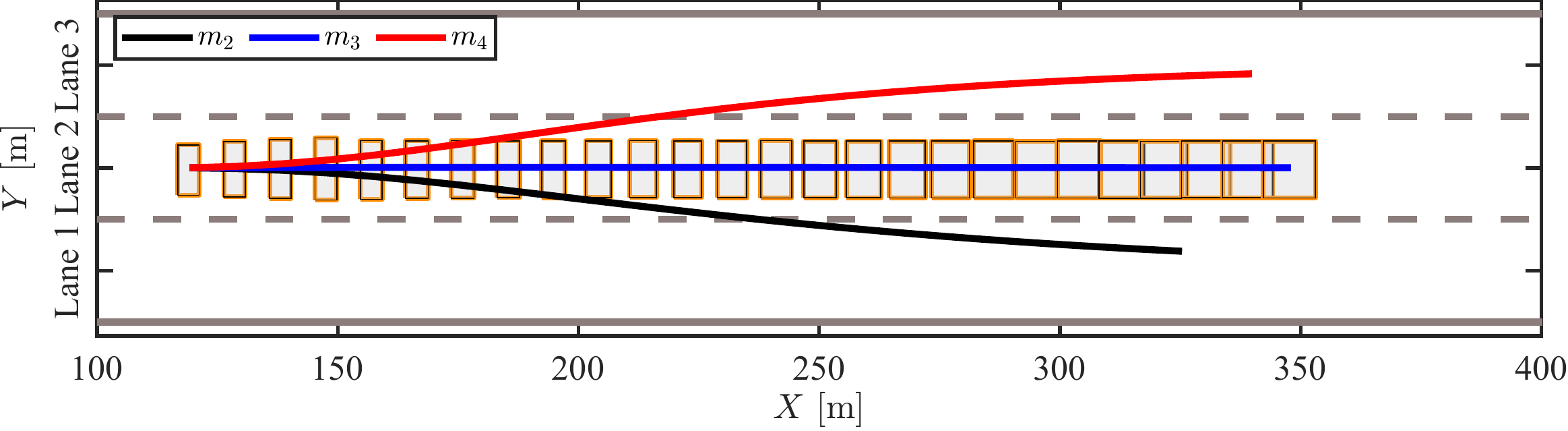}%
		\label{Fig_GMM_obst_road__b}}
	\caption{The safety-aware occupancy over the prediction horizon. (a) Parameter $\varepsilon_k = 0.2$. (b) Parameter $\varepsilon_k = 0.8$.}
	\label{Fig_GMM_obst_road}
\end{figure}

In Fig.~\ref{Fig_GMM_obstacle}(a), three maneuvers are involved in $\mathcal{O}_t^c$, while in Fig.~\ref{Fig_GMM_obstacle}(b) only the maneuver $m_3$ is considered in $\mathcal{O}_t^c$ since $\varepsilon_k = 0.8$ only intersects with the highest peak in Fig.~\ref{Fig_intercept_GMM}. Therefore, reducing the value of $\varepsilon_k$ will clearly formulate a more conservative set $\mathcal{O}_t^c$ based on the GMM-distribution assumption. Further, the vehicle shape (vehicle length $l_{\rm veh}$ and width $w_{\rm veh}$) are considered to expand the occupancy $\mathcal{O}_t^c$:
\begin{subequations}
	\begin{align}
		\mathcal{O}_{{\rm veh},t}^c & = \mathcal{O}_t^c \oplus S_{\rm veh} \label{eq_gmm_obstacle_vehicle_shape_a} \\
		\begin{split}
			S_{\rm veh} & = \left\{ (x, y)\bigg\vert{\begin{matrix} -\frac{l_{\rm veh}}{2} \leq x \leq \frac{l_{\rm veh}}{2} \\  -\frac{w_{\rm veh}}{2} \leq y \leq \frac{w_{\rm veh}}{2} \end{matrix}} \right\} \label{eq_gmm_obstacle_vehicle_shape_b}
		\end{split}
	\end{align}
\end{subequations}
\noindent where for simplification, it is assumed that all vehicles have the same length $l_{\rm veh}$ and width $w_{\rm veh}$. The impact of the heading angle of the SV $c$ is ignored in \eqref{eq_gmm_obstacle_vehicle_shape_a}.

The set $\mathcal{O}_{{\rm veh}, t}^c$ is called the safety-aware occupancy because it is explicitly related to the safety-awareness parameter $\varepsilon_k$. A smaller $\varepsilon_k$ enlarges the obstacle's occupancy such that it cares more about the safety/robustness of the motion planning, while a larger $\varepsilon_k$ defines a smaller occupancy of the obstacle, which increases the feasible region of the motion-planning algorithm under the premise of potentially reducing safety. The length of $\mathcal{O}_{{\rm veh}, t}^c$ is denoted as $L_t^c$ and the width is $W_t^c$, and the longitudinal and lateral centers are denoted as $o_{x, t}^c$ and $o_{y,t}^c$, respectively. The visualization of $\mathcal{O}_{{\rm veh}, t}^c$ over the prediction horizon is shown in Fig.~\ref{Fig_GMM_obst_road}, where the probabilities of $m_2, \ m_3, \ m_4$ are 0.14, 0.6, and 0.26, respectively. The gray areas indicate  $\mathcal{O}_{{\rm veh}, t}^c$ over the prediction horizon.

\begin{remark}
	There are different possible shapes to construct $\mathcal{O}_t^c$, like a convex hull \cite{cesari2017scenario}, an ellipse \cite{brudigam2021collision}, or a polytope  \cite{ahn2021safe}. This paper uses a simple rectangle because it is suitable for constructing the linear safety-distance constraints in the MPC. The details will be discussed in Section~\ref{MPC OCP}.	
\end{remark}

\subsection{Computing Moving Targets for the MPC}\label{Moving Target}
Moving targets, which are defined as $\tilde{v}^{\rm EV}_{x, k} $ and $\tilde{y}^{\rm EV}_k$ in \eqref{eq_ocp_a}, are the references for the MPC at time step $k$, where the iteration step $k$ indicates that the references are time-varying. In this paper, the approach in \cite{zhou2022interaction} that computes the reference targets in a deterministic MPC is extended to compute $\tilde{v}^{\rm EV}_{x, k} $ and $\tilde{y}^{\rm EV}_k$ based on the safety-aware occupancy of the SVs. 

Let the EV's maneuver set $\mathcal{M}_{\rm EV}$ have three candidate actions, namely, velocity-tracking maneuver on lane~1 (${\rm VT1}$), lane~2 (${\rm VT2}$), lane~3 (${\rm VT3}$), such that $\mathcal{M}_{\rm EV} = \left\{{\rm VT1}, {\rm VT2}, {\rm VT3}\right\}$. For each maneuver $m \in \mathcal{M}_{\rm EV}$, the lateral reference position $\tilde{y}^{{\rm EV},m}_k$ is the center line of the corresponding lane, while the longitudinal reference velocity $\tilde{v}^{{\rm EV},m}_{x, k}$ is collision-free with respect to SVs. Define the binary parameter $z_t \in \left\{0, 1\right\}$. Then $\tilde{v}_{x, k}^{{\rm EV}, m}$ can be computed by solving a quadratic programming (QP) problem:
\begin{subequations}
	\begin{align}
		\mathop{\rm minimize}\limits_{\tilde{v}_{x, k}^{{\rm EV}, m}} \ \ & \left\|v_{x, k}^{\rm EV} - \tilde{v}_{x, k}^{{\rm EV}, m}\right\|_2^2 & \label{eq_mt_velocity_a}\\
		{\rm subject\ to} \ \
		& z_t\cdot (x_{t|k}^{{\rm EV}, m} + d_s) \leq z_t\cdot o_{x, t}^c, {\forall}c \in \mathcal{C}_k &\label{eq_mt_velocity_b}
	\end{align}
\end{subequations}
\noindent where $v_{x, k}^{\rm EV}$ is the speed of the EV at time step $k$ in the ground coordinate system, and $d_s = \tilde{v}_{x, k}^{{\rm EV},m} \cdot \tau_h + (L_t^c + l_{\rm veh})/2$ defines the safety distance by the time headway $\tau_h$. The parameter $z_t$ is defined as:
\begin{equation}
	z_t = \begin{cases}
		1 &  {{\rm if} \ \left|y_{t|k}^{{\rm EV},m} - o_{y, t}^c\right| \leq \frac{w_{\rm veh} + W_t^c}{2}} \\ 
		0 &  {\rm otherwise}
	\end{cases}
	\label{eq_lateral_close}
\end{equation}
where $z_t = 1$ means that the EV and the occupancy of SV $c$ are laterally close to each other. In \eqref{eq_mt_velocity_b} and \eqref{eq_lateral_close}, $x_{t|k}^{{\rm EV},m}$ and $y_{t|k}^{{\rm EV},m}$ mean the predicted longitudinal and lateral positions of maneuver $m$ of the EV at time step $t$, and they are predicted by the linear state-feedback model \eqref{eq_ls_a}--\eqref{eq_ls_c} based on some adaptations. Specifically, the controller gains $\hat{K}_{\rm lon}^m$ and $\hat{K}_{\rm lat}^m$ in \eqref{eq_ls_a} are replaced by the longitudinal controller gains $K_{\rm lon}^{\rm EV}$ and lateral controller gains $K_{\rm lat}^{\rm EV}$ of the EV. In addition, the references $\tilde{v}_{x, k}^{c, m}$ and $\tilde{y}_k^{c, m}$ in \eqref{eq_ls_c} are replaced by $\tilde{v}^{{\rm EV},m}_{x, k} $ and $\tilde{y}^{{\rm EV},m}_k$, respectively. The state $X_k^{c, m}$ in \eqref{eq_ls_a} is replaced by the measured state of the EV in the ground coordinate system.

After computing the target speed of every maneuver $m$ for the EV, the maneuver cost can be evaluated by \cite{zhou2020multiobjective}:
\begin{equation}
	\begin{split}
		J_k^m = & \sum_{t=k}^{k+N}\left( \left\| a_{x, t|k}^{{\rm EV},m} \right\|_{W_x}^2 + \left\|a_{y,t|k}^{{\rm EV},m}\right\|_{W_y}^2\right) + \\ &
		\left\|v_{x, k}^{{\rm EV}} - \tilde{v}_{x, k}^{{\rm EV}, m}\right\|_{W_v}^2 + 
		\left\|y_{k}^{{\rm EV}} - \tilde{y}_k^{{\rm EV}, m}\right\|_{W_l}^2 \label{eq_model_cost}
	\end{split}
\end{equation} 
\noindent where $W_x$, $W_y$, $W_v$, and $W_l$ are weights, $a_{x, t|k}^{{\rm EV},m}$ and $a_{y, t|k}^{{\rm EV},m}$ are the longitudinal and lateral accelerations of the EV predicted by the adapted linear state-feedback model \eqref{eq_ls_a}, and $y_{k}^{\rm EV}$ is the lateral position of the EV at time step {\it k}. The cost is used to compute the probability $p_k^{{\rm EV}, m}$ of maneuver $m$ \cite{lefkopoulos2021interaction}:
\begin{equation}
		p_k^{{\rm EV}, m} =  \frac{1/\sqrt{J_k^m + \varsigma}}{\sum_{q \in \mathcal{M}_{\rm EV}}1/\sqrt{J_k^q + \varsigma}} \label{eq_model_pro}
\end{equation}
\noindent where $\varsigma$ is a small positive constant in case the maneuver cost is zero. Then, the EV's optimal maneuver $m_k^{\rm EV}$ is chosen by:
\begin{equation}
	m_k^{\rm EV} = {\rm argmax_{\it m \in \mathcal{M}_{\rm EV}}} \ \left\{p_k^{{\rm EV}, m}\right\} \label{eq_select_model}
\end{equation}

The reference targets associated with the optimal maneuver $m_k^{\rm EV}$ are the moving targets for time step $k$.

\subsection{Extracting the Most Direct Obstacles}\label{Most Direct Vehicle}
\begin{algorithm}[!t]\scriptsize
	\DontPrintSemicolon
	\KwIn{$m_k^{\rm EV}$, $y_k^{\rm EV}$, $\phi_k^{\rm EV}$, $\mathcal{O}_{{\rm veh}, t}^c$, $\forall c \in \mathcal{C}_k$, $t=k+1, \ldots, k+N$.}
	\KwOut{$x_t^{\rm DV}$: Longitudinal position of DV at time step $t$.}
	Compute $\left\lceil{\frac{y_k^{\rm EV}-\zeta_{\rm EV}}{w_{\rm lane}}}\right \rceil = C_L$, $x_t^c = o_{x, t}^c - \frac{L_{t}^c}{2}$. \\
	Define parameters $z_t^{c, 1} = +\infty$, $z_t^{c, 2} = +\infty$, $z_t^{c, 3} = +\infty$. \\
	\If{$\left(\left\lceil{\frac{o_{y, t}^c - W_t^c/2}{w_{\rm lane}}}\right \rceil = 1 \ \bigg| \ \left\lceil{\frac{o_{y, t}^c}{w_{\rm lane}}}\right \rceil = 1\right)$}
	{let $z_t^{c, 1} = 1$}
	\If{$\left(\left\lceil{\frac{o_{y, t}^c -  W_t^c/2}{w_{\rm lane}}}\right \rceil = 2 \ \bigg| \ \left\lceil{\frac{o_{y, t}^c}{w_{\rm lane}}}\right \rceil = 2 \ \bigg| \ \left\lceil{\frac{o_{y, t}^c +  W_t^c/2}{w_{\rm lane}}}\right \rceil = 2\right)$}
	{let $z_t^{c, 2} = 1$}
	\If{ $\left(\left\lceil{\frac{o_{y, t}^c}{w_{\rm lane}}}\right \rceil = 3 \ \bigg| \ \left\lceil{\frac{o_{y, t}^c +  W_t^c/2}{w_{\rm lane}}}\right \rceil = 3\right)$}
	{let $z_t^{c, 3} = 1$}
	\If{$C_L = 1$}
	{
		$d_y = y_k^{\rm EV} - 0.5w_{\rm lane}$ \\
		\If{$\left\{(m_k^{\rm EV} = {\rm VT2}) \ \Big| \ (\phi_k^{\rm EV} \geq \lambda_{\phi} \ \& \ d_y \geq \lambda_{y})\right\}$}
		{$x_t^{\rm DV} = {\rm minimize}\left\{ z_t^{c, 1} \cdot x_t^{\rm c}, \ z_t^{c, 2} \cdot x_t^{\rm c} \right\}$}
		\Else{$x_t^{\rm DV} = {\rm minimize}\left\{ z_t^{c, 1} \cdot x_t^{\rm c}\right\}$}
	}
	\ElseIf{$C_L = 2$}
	{$d_y = y_k^{\rm EV} - 1.5w_{\rm lane}$ \\
		\If{$\left\{(m_k^{\rm EV} = {\rm VT1}) \Big| (\phi_k^{\rm EV} \leq -\lambda_{\phi} \ \& \ d_y \leq -\lambda_{y})\right\}$}
		{$x_t^{\rm DV} = {\rm minimize}\left\{ z_t^{c, 1} \cdot x_t^{\rm c}, \ z_t^{c, 2} \cdot x_t^{\rm c} \right\}$}
		\ElseIf{$\left\{(m_k^{\rm EV} = {\rm VT3}) \Big| (\phi_k^{\rm EV} \geq \lambda_{\phi} \ \& \ d_y \geq \lambda_{y})\right\}$}
		{$x_t^{\rm DV}={\rm minimize}\left\{ z_t^{c, 2} \cdot x_t^{\rm c}, \ z_t^{c, 3} \cdot x_t^{\rm c} \right\}$}
		\Else{$x_t^{\rm DV} = {\rm minimize}\left\{ z_t^{c, 2} \cdot x_t^{\rm c}  \right\}$}
	}
	\Else
{
	$d_y = y_k^{\rm EV} - 2.5w_{\rm lane}$ \\
	\If{$\left\{(m_k^{\rm EV} = {\rm VT2}) \ \Big| \ (\phi_k^{\rm EV} \leq -\lambda_{\phi} \ \& \ d_y \leq -\lambda_{y})\right\}$}
	{$x_t^{\rm DV}={\rm minimize}\left\{ z_t^{c, 2} \cdot x_t^{\rm c}, \ z_t^{c, 3} \cdot x_t^{\rm c}\right\}$}
	\Else{$x_t^{\rm DV} = {\rm minimize}\left\{ z_t^{c, 3} \cdot x_t^{\rm c}\right\}$}
}
	\caption{Extracting DVs from multiple SVs}
	\label{alg:DV}
\end{algorithm}

In multi-vehicle scenarios, the obstacle occupancy is computed at every time step over the prediction horizon. In order to reduce the scale of the optimization problem, it is beneficial to extract the most direct vehicles (DVs) from multi-vehicle situations to construct the safety constraint in the MPC. This is achieved by Algorithm~\ref{alg:DV}, where $\zeta_{\rm EV}$ is a margin of $y_k^{\rm EV}$; $\lambda_{\phi}, \ \lambda_{y}$ are thresholds for $\phi_k^{\rm EV}$ and $d_y$, and $w_{\rm lane}$ is the lane width. Here, the second scenario (line 15--line 22) where $C_L = 2$ (the EV is in lane 2 at time step $k$) is interpreted:
\begin{itemize}
\item Line 17 means that the intended maneuver of the EV is ${\rm VT1}$ (i.e., the EV wants to go to lane~1), or both $d_y$, i.e., the deviation between the EV's current position $y_k^{\rm EV}$ and the current lane center line ($1.5w_{\rm lane}$), and $\phi_k^{\rm EV}$ are beyond their thresholds, such that the EV has a possibility to go to lane~3 over the horizon. In this case, the occupancy of the SVs that may appear in both lane~1 and lane~2 are considered. Then line 18 is applied to find the closest occupancy for time step $t$.

\item Line 19 means that EV intends to go to lane~3, or it has a possibility to go to lane~3 because of its state at time step $k$. Then, line~20 is applied to find the closest occupancy of the SVs, which may appear in both lane~2 and lane~3 over the prediction horizon.
	
\item  Line~21 means that the EV intends to keep on lane~2, and will not go to lane~1 or lane~3 over the prediction horizon. Hence, only the occupancies of the SVs that may appear in lane~2 are considered to define the DV on line~22.  
\end{itemize}
\begin{figure}[!t]
	\includegraphics[width = \columnwidth]{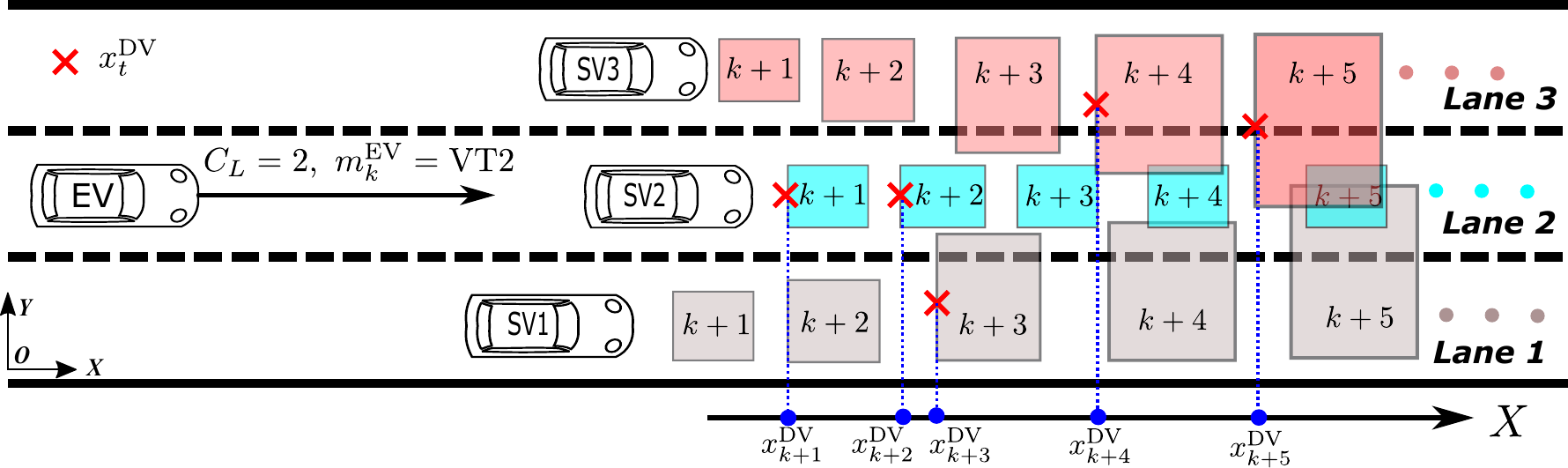}
	\caption{Example of extracting the DVs over the prediction horizon.}
	\label{Fig_DV}
\end{figure}

To illustrate the idea of Algorithm~\ref{alg:DV}, the situation of line 21--22 is visualized by an example shown in Fig.~\ref{Fig_DV}. This example illustrates that if the EV's intended maneuver is ${\rm VT2}$, and it will keep in lane~2 over the prediction horizon, then the DV at every time step over the prediction horizon is selected as the closest occupancy of each SV that expands to lane 2. The conditions of the other lines can be inferred similarly.

\subsection{Designing the MPC Controller}\label{MPC OCP}
Based on the references and the DV, the OCP problem of the MPC that has been formulated in \eqref{eq_ocp_a}--\eqref{eq_ocp_f} is specified as:
\begin{subequations}
	\begin{align}
		\mathop{\rm minimize}\limits_{U_{t-1}, \ \rho} \quad & \sum_{t = k+1}^{k+N}\left(\left\|s^{\rm EV}_{t-1}\right\|_{Q_1}^2 +
		\left\| \alpha^{\rm EV}_{t-1}\right\|_{Q_2}^2\right) \notag \\ &
		+ \sum_{t = k+1}^{k+N}\left(\left\|a^{\rm EV}_{t|k}\right\|_{Q_3}^2 + \left\| \delta^{\rm EV}_{t|k}\right\|_{Q_4}^2\right) \notag \\ & + \left\|E_{k+N|k}\right\|_{Q_5}^2 + \left\|\rho\right\|_{Q_6}^2 & \label{eq_spe_ocp_a} \\
		{\rm subject\ to}\quad
		&\xi^{\rm EV}_{t|k} = f(\xi^{\rm EV}_{t-1|k}, \ U^{\rm EV}_{t-1}) & \label{eq_spe_ocp_b}\\
		&\underline{\mathcal{U}} \leq \left[\delta^{\rm EV}_{t|k} \ a^{\rm EV}_{t|k} \ v^{\rm EV}_{t|k} \right]^{\rm T} \leq \overline{\mathcal{U}} &\label{eq_spe_ocp_c}\\
		&v^{\rm EV}_{t|k}\tau_h-\rho \leq x_t^{\rm DV} - x^{\rm EV}_{t|k} - \frac{l_{\rm veh}}{2}&\label{eq_spe_ocp_d}\\
		&\rho \in \mathcal{R}_+^N, \ t = k+1,\dots,k+N& \label{eq_spe_ocp_e}
	\end{align}
\end{subequations}
\noindent In the cost function, $E_{k+N|k}$ is the terminal deviation vector, i.e., $E_{k+N|k} = \left[y^{\rm EV}_{k+N|k}-\tilde{y}_k^{\rm EV} \ v^{\rm EV}_{k+N|k}- \tilde{v}_{x, k}^{\rm EV}\right]$, where $y^{\rm EV}_{k+N|k}$ and $v^{\rm EV}_{k+N|k}$ are the terminal lateral position and longitudinal speed, respectively, predicted by the EV model~\eqref{eq_vehicle_model_ode}. The loss function weighting matrices are denoted by $Q_1, \ldots, Q_6$. The constraint~\eqref{eq_spe_ocp_b} enforces the vehicle model~\eqref{eq_vehicle_model_ode}, and constraint~\eqref{eq_spe_ocp_c} defines the bounds on the vehicle states. The constraint~\eqref{eq_spe_ocp_d} enforces the safety distance between the EV and DVs based on the time headway $\tau_h$. The constraint~\eqref{eq_spe_ocp_e} applies to the slack variables $\rho$. The proposed MPC-based motion-planning method achieves interaction awareness through the interaction-aware motion prediction of the IAIMM-KF, and safety awareness through tuning the parameter $\varepsilon_k$ in \eqref{eq_intercept_pdf}. It is therefore called interaction-safety-aware MPC (ISA-MPC). 
\begin{remark}
	It is worth noting that the policy-based approach \cite{batkovic2021robust}, \cite{nair2022stochastic}, as mentioned in Section~\ref{related work planning}, has shown good performance in reducing the conservatism compared to the sequence-based approach in solving problems with multi-modal obstacle uncertainties. A performed numerical study\footnote{The implementation and results of the numerical study can be found together with the published code.} revealed that the policy-based approach exhibits reduced conservatism, particularly when the probability of one of the modes is predicted to be very small, while the solution time is also much longer.  However, it is important to emphasize that the policy-based approach can be costly for solving motion-planning problems in multi-vehicle scenarios. This is primarily because of three factors. Firstly, the approach requires solving multiple OCPs at each time step, resulting in increased computational demands. Secondly, the number of policies increases exponentially with the number of SVs and their modes. It is possible to restrict the number of policies by considering the most probable 2--3 maneuvers of the SVs\cite{tas2018decision}, while the computational effort remains a challenge in multi-vehicle scenarios. Thirdly, the policy-based approach necessitates reformulating the OCP at every time step, as the number of optimization variables of the OCP, which depend on the prediction of obstacle motion, is changing (see Eq. (7) in \cite{batkovic2021robust}\footnote{An example is provided together with the code to illustrate how the prediction of obstacles decides the number of optimization variables.}). This reformulation process becomes more complex with multiple SVs and multi-modal uncertainties, further adding to the computational burden. In contrast, the OCP \eqref{eq_spe_ocp_a}--\eqref{eq_spe_ocp_e} maintains a fixed structure, ensuring that the solution time remains unaffected by the traffic environment (see Table~\ref{Table_computation_time}). Furthermore, the parameter $\varepsilon_k$ can be tuned to adjust the conservatism of the sequence-based optimization (see Figs.~\ref{Fig_SIM_1_GMMMPC} and \ref{Fig_SIM_3}). These characteristics make the proposed MPC efficient and implementable in multi-vehicle scenarios.
\end{remark}

\section{Results and Discussion}\label{Simulations and Discussions}
The parameters used for the simulations are collected in Table~\ref{Table_simulation_parameters}. The details regarding the key implementation aspects of the method are highlighted here:
\begin{itemize}
	\item The NLS problem \eqref{eq_nls_single_a}--\eqref{eq_nls_single_e} is solved offline by the MATLAB solver \texttt{lsqnonlin} with the Levenberg--Marquardt algorithm~\cite{nocedal1999numerical} (results are provided in the code).
	\item The OCP problem in \eqref{eq_spe_ocp_a}--\eqref{eq_spe_ocp_e}
	is solved online by \texttt{CasADi} and \texttt{Ipopt} using the linear solver \texttt{MA57} \cite{andersson2019casadi, wachter2006implementation, hsl2021collection} with a warm start as the solution of the previous problem. This is achieved by activating the \texttt{Ipopt} option \texttt{warm\_start\_init\_point}. The \texttt{MA57} solver is chosen based on the conclusions in \cite{tasseff2019exploring}.
	\item  In \eqref{eq_intercept_pdf}, the set $\mathcal{D}_t^c$ does not have an analytical expression. Therefore, in order to determine this set numerically, the Python package \texttt{skimage} \cite{van2014scikit} is utilized.
\end{itemize}

The simulations are conducted in three cases. In Case I, the proposed ISA-MPC is compared with a scenario MPC (SCMPC) approach to investigate the impact of the risk-aware obstacle occupancy of the ISA-MPC and the sampling-based obstacle occupancy of the SCMPC on the performance of the planned trajectory. In the second case, the ISA-MPC is compared with a deterministic MPC, which was used in \cite{zhou2022interaction}, to verify its performance in the presence of multi-modal obstacle uncertainties. In the final case, the ISA-MPC is compared with a human-driven vehicle from a driving scenario of the highD dataset to evaluate the performance of the method on the real-world traffic dataset. Simulation videos are accessible online\footnote{\url{https://youtu.be/7we4Vb4zftY}}.
\begin{table}[!t]\scriptsize
	\centering
	\caption{Simulation parameters}  
	\label{Table_simulation_parameters} 
\begin{tabular}{cccc}
	\toprule
	\textbf{Symbol} & \textbf{Value} & \textbf{Symbol} & \textbf{Value} \\
	\midrule
	$K_{\rm lon}^{\rm EV}$ & $[0.1029 \ 0.3423]$ & $N$& $25$ \\ \specialrule{0em}{1pt}{1pt}
	$K_{\rm lat}^{\rm EV}$ & $[0.0984 \ 0.4656 \ 0.5417]$ & $T$ & $0.32 \ {\rm s}$ \\ \specialrule{0em}{1pt}{1pt}
	$W_x, W_y, W_v, W_l$ & $0.1, \ 0.3, \ 0.1, \ 0.5$ & $\varsigma$& $10^{-4}$\\ \specialrule{0em}{1pt}{1pt}
	$l_f, l_r$ &  $1.477 \ {\rm m}$, $1.446 \ {\rm m}$ & $w_{\rm lane}$ & $3.75 \ {\rm m}$\\ \specialrule{0em}{1pt}{1pt}
	$\underline{\mathcal{U}}$& $[-0.8 \ -6 \ 0]^{\rm T}$ & $K_{\rm sam}$ & $30$\\ \specialrule{0em}{1pt}{1pt}
	$\overline{\mathcal{U}}$ & $[0.8  \ 6  \ +\infty]^{\rm T}$ & $\zeta_{\rm EV}$& $0.5 \ {\rm m}$\\ \specialrule{0em}{1pt}{1pt}
	$Q_1, Q_3$ & 0.5 & $Q_2, Q_4$ & 0.1 \\ \specialrule{0em}{1pt}{1pt}
	$Q_5$ & $\rm{diag}([0.05 \quad 1])$ & $Q_6$ & $0.055\cdot I^N$ \\ \specialrule{0em}{1pt}{1pt}
	$w_{\rm vel}, l_{\rm vel}$ & $1.8 \ {\rm m}$, $4.3 \ {\rm m}$ &$\tau_h$ &  2 ${\rm s}$ in \eqref{eq_spe_ocp_d} \\ \specialrule{0em}{1pt}{1pt}
	$\lambda_{\phi}, \lambda_y$ &  $0.015 \ {\rm rad}$, $1.8  \ {\rm m}$ & $\tau_h$ & $1.5 \ {\rm s}$ in \eqref{eq_mt_velocity_b} \\ \specialrule{0em}{1pt}{0.1pt}
	\bottomrule
\end{tabular}
\end{table}

\subsection{{\rm Case I}: ISA-MPC vs. SCMPC}\label{SIM1}
The SCMPC is an alternative way to consider multi-modal uncertainties of SVs \cite{cesari2017scenario}, \cite{brudigam2018combining}. In this simulation, the proposed ISA-MPC will be compared with a sequence optimization-based SCMPC, which uses the sampling-based approach to formulate the obstacle occupancy such that the problem has the same computational complexity as the OCP \eqref{eq_spe_ocp_a}--\eqref{eq_spe_ocp_e}. In order to adapt the scenario approach to the MPC formulation of this paper, we did not directly compare with the SCMPC approaches in \cite{cesari2017scenario} and \cite{brudigam2018combining} which fixed the reference of the MPC at every time step. In this paper, we sampled the maneuvers of the SVs according to the maneuver probabilities to formulate the occupancy of SVs like in Fig.~\ref{Fig_GMM_obst_road} to compute the moving targets as in Section~\ref{Moving Target} and to formulate the OCP as in Section~\ref{MPC OCP}. The idea of quantifying SVs' occupancy in the comparison SCMPC is inspired by the scenario-generation methods in \cite[Section IV-A]{cesari2017scenario} and \cite[Section II-B]{brudigam2018combining}. Then, the obstacle occupancies formulated by the scenario approach and the safety-aware approach, respectively, are substituted into the OCP \eqref{eq_spe_ocp_a}--\eqref{eq_spe_ocp_e} for comparison.

\begin{figure}[!t]
	\centering
	\includegraphics[width = \columnwidth]{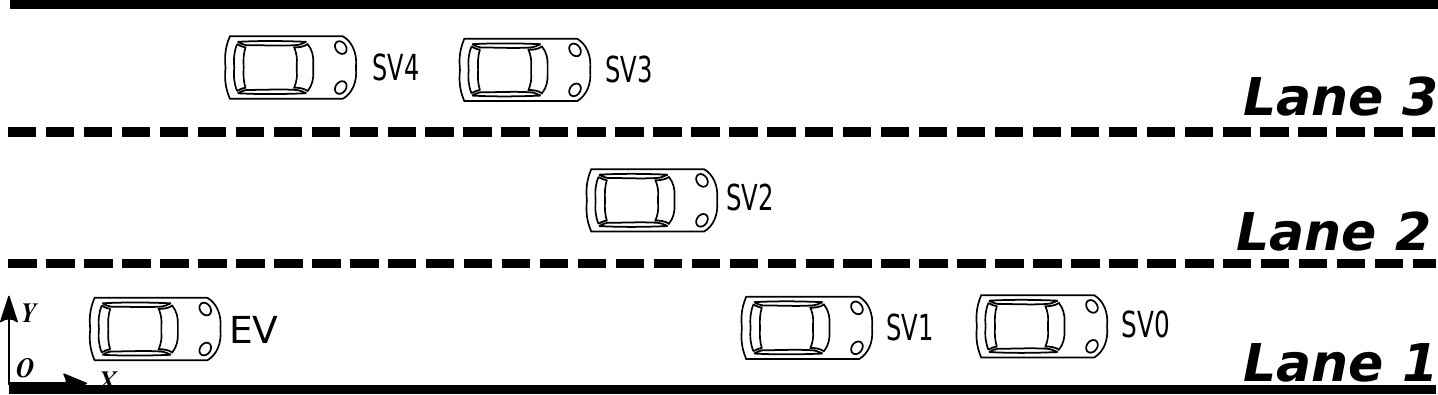}
	\caption{Simulation scenario for Section~\ref{SIM1} and Section~\ref{SIM2}.}
	\label{Fig_SIM_description}
\end{figure}

\begin{figure}[!t]
	\centering
	\includegraphics[width = \columnwidth]{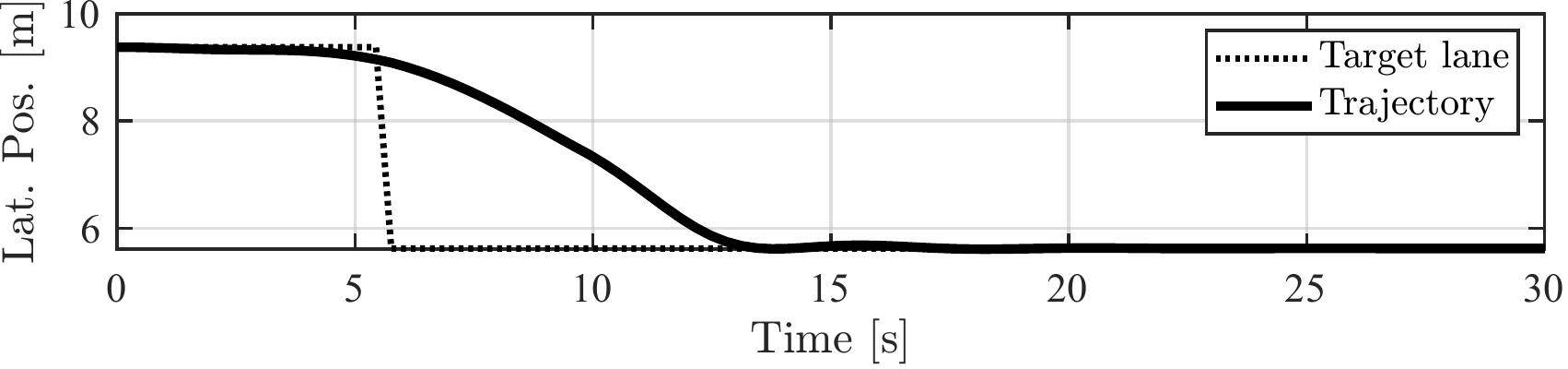}
	\caption{The cut-in trajectory of SV4.}
	\label{Fig_SIM1_GMMMPC_SV4}
\end{figure}

The simulations are conducted on a three-lane road, as shown in Fig.~\ref{Fig_SIM_description}, where the nominal speeds of lane~1 to lane~3 are $65 \ {\rm km/h}$, $90 \ {\rm km/h}$, and $90 \ {\rm km/h}$, respectively. The initial longitudinal positions and speeds of the vehicles are: SV0 ($200 \ {\rm m}$, $60 \ {\rm km/h}$), SV1 ($150 \ {\rm m}$, $60 \ {\rm km/h}$), SV2 ($100 \ {\rm m}$, $108 \ {\rm km/h}$), SV3 ($85 \ {\rm m}$, $95 \ {\rm km/h}$),  SV4 ($35 \ {\rm m}$, $95 \ {\rm km/h}$), EV ($0 \ {\rm m}$, $105 \ {\rm km/h}$), with zero initial accelerations. According to the approach in Section~\ref{Moving Target}, the optimal maneuver of the EV at the beginning is to change lanes, and other vehicles are simulated by the IAIMM-KF to keep their lanes. In order to trigger a challenging scenario for the EV, let SV3 brake suddenly after time step 3 ($0.96 \ {\rm s}$) with a constant deceleration of ${\rm -1.2~m/s^2}$. Then SV4 needs to change lanes to avoid a collision with SV3, see Fig.~\ref{Fig_SIM1_GMMMPC_SV4}. This maneuver happens in the lane-changing process of the EV, so SV4 abruptly becomes a cut-in obstacle for the EV. 

The ISA-MPC and SCMPC are applied separately to plan the motion for the EV in the same situation. In ISA-MPC, the safety-awareness parameter $\varepsilon_k$ is set as 0.8, 0.5, and 0.2, respectively. In SCMPC, the corresponding tuning parameter is the sampling size $K_{\rm sc}$, which is set to 5, 10, and 15, respectively. These simulations are to compare how the formulation of the obstacle occupancy affects the performance of the MPC. For simplification, the nominal trajectories and probabilities of the maneuvers of SVs simulated by the IAIMM-KF are directly substituted into the MPC of the EV, while the STDs of positions are estimated according to the approach in Section~\ref{Estimating Trajectory Uncertainty} to formulate the obstacle occupancy. 

\begin{figure}[!t]
	\centering
	\subfloat[ ]{\includegraphics[width = \columnwidth]{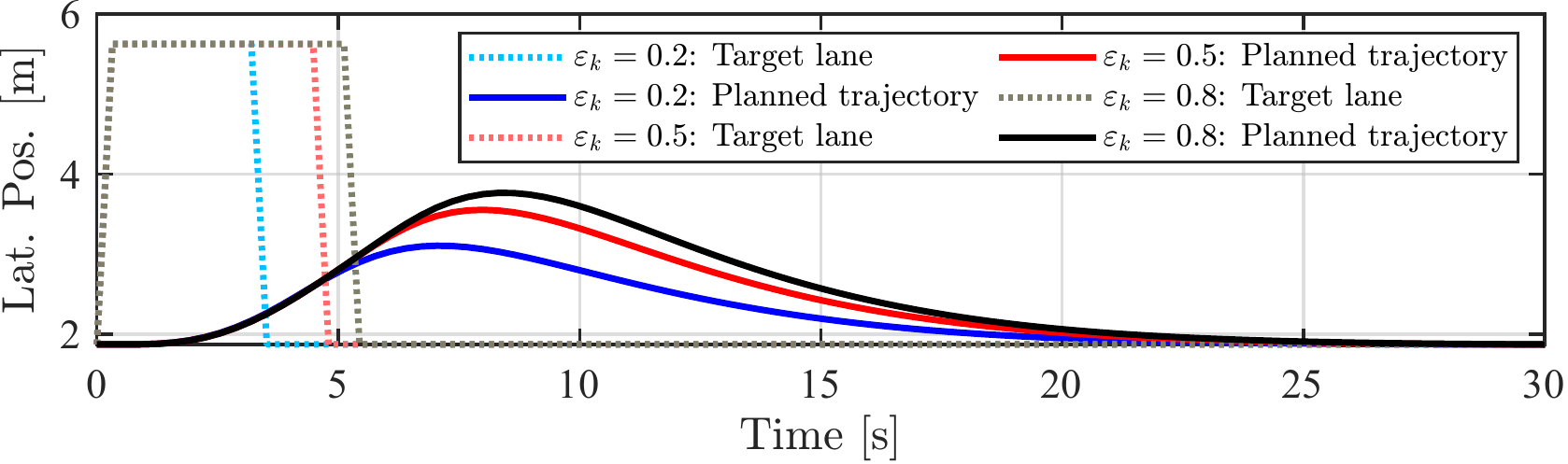}%
		\label{Fig_SIM1_GMMMPC_y}} 
	\vspace{0 cm}
	\subfloat[ ]{\includegraphics[width = \columnwidth]{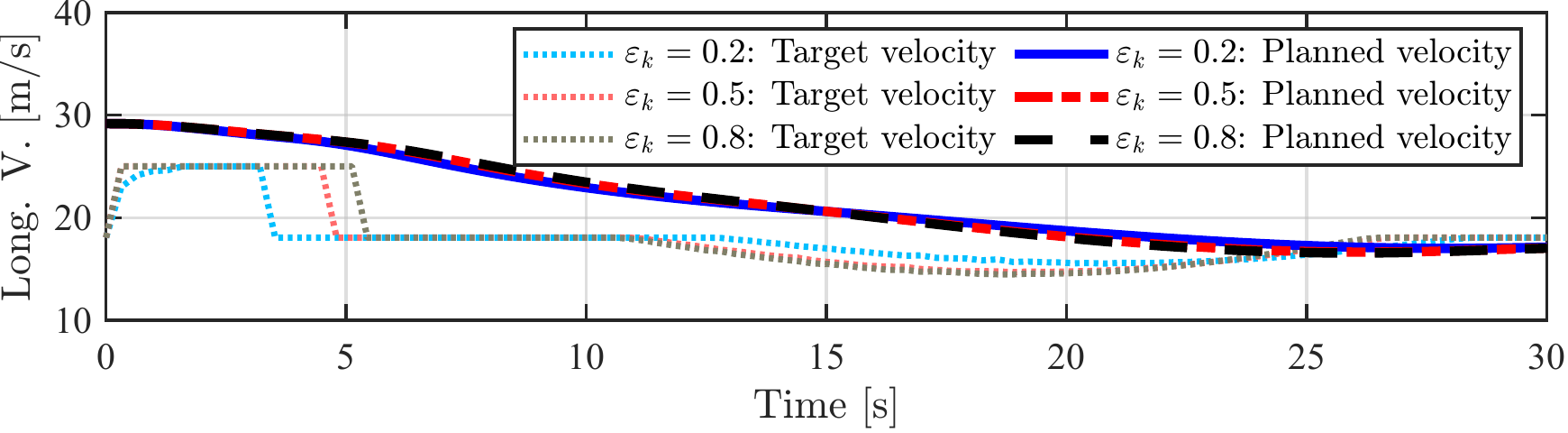}%
		\label{Fig_SIM1_GMMMPC_v}}
	\caption{Planned trajectories of the EV by ISA-MPC with different values of $\varepsilon_k$. (a) Lateral position of the EV. (b) Longitudinal velocity of the EV.}
	\label{Fig_SIM_1_GMMMPC}
\end{figure}

\begin{figure}[!t]
	\centering
	\subfloat[ ]{\includegraphics[width = \columnwidth]{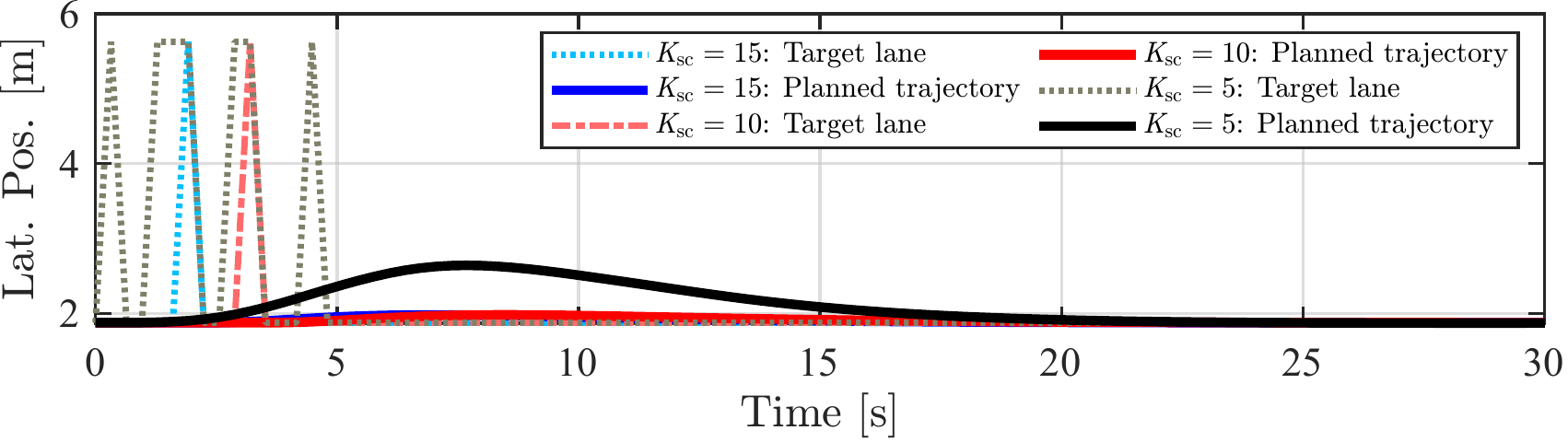}%
		\label{Fig_SIM1_SCMPC_y}} 
	\vspace{0 cm}
	\subfloat[ ]{\includegraphics[width = \columnwidth]{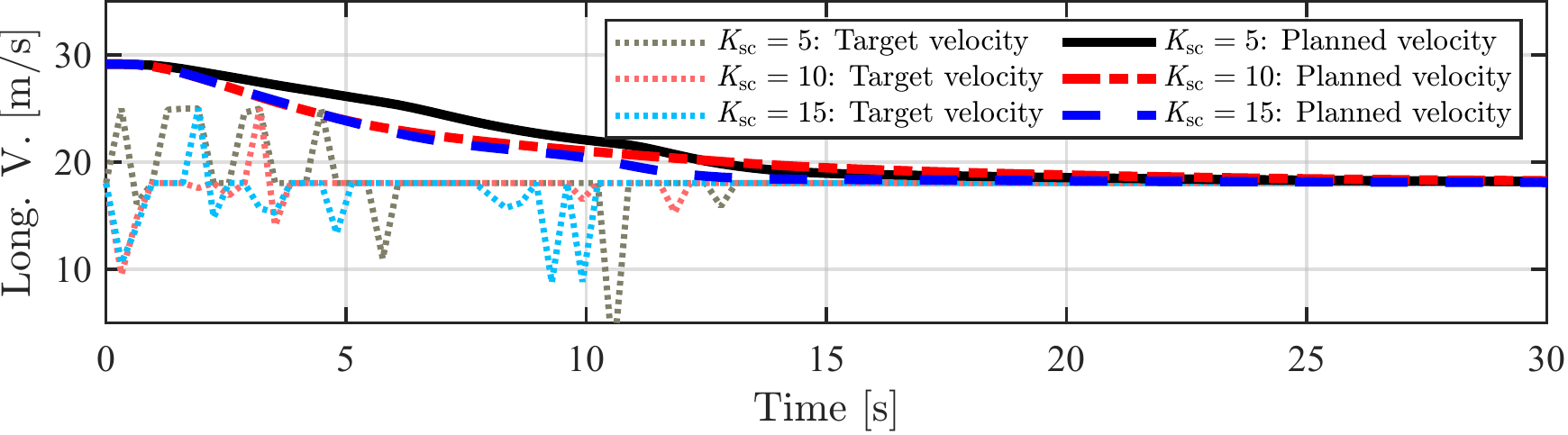}%
		\label{Fig_SIM1_SCMPC_v}}
	\caption{Planned trajectories of the EV by SCMPC with different values of $K_{\rm sc}$. (a) Lateral position. (b) Longitudinal velocity.}
	\label{Fig_SIM_1_SCMPC}
\end{figure}

The simulation results are shown in Fig.~{\ref{Fig_SIM_1_GMMMPC}} and Fig.~{\ref{Fig_SIM_1_SCMPC}}. Fig.~\ref{Fig_SIM_1_GMMMPC} shows that ISA-MPC is proactive because the EV considers the probability of a lane-changing maneuver of the SV4 before the SV decides to change lane. In addition, a smaller $\varepsilon_k$ makes the planner more proactive, while a larger $\varepsilon_k$ makes the EV  respond slower. For the SCMPC, a larger $K_{\rm sc}$ should make the planner more conservative, while a smaller $K_{\rm sc}$ makes the planner more aggressive \cite{schildbach2014scenario}. However, Fig.~\ref{Fig_SIM_1_SCMPC} shows that the EV frequently updates the targets because of the randomness in sampling from the multi-modal distribution of the motion of SVs. This process is clearly observed in the simulation videos. 

Comparisons between Fig.~{\ref{Fig_SIM_1_GMMMPC}} and Fig.~{\ref{Fig_SIM_1_SCMPC}} show that the performance of ISA-MPC is well-tuned by the parameter $\varepsilon_k$ that makes the method certainly becoming more proactive when decreasing its value, while the randomness of SCMPC does not guarantee that the planner always performs as desired.

\subsection{{\rm Case II}: ISA-MPC vs. Deterministic MPC}\label{SIM2}

This simulation compares the ISA-MPC and the deterministic MPC with the simulation scenario the same as in Fig.~\ref{Fig_SIM_description}. Here, the longitudinal position and velocity of the SVs in Fig.~\ref{Fig_SIM_description} are initialized as SV0 ($250 \ {\rm m}$, $60 \ {\rm km/h}$), SV1 ($200 \ {\rm m}$, $60 \ {\rm km/h}$), SV2 ($100 \ {\rm m}$, $108 \ {\rm km/h}$), SV3 ($80 \ {\rm m}$, $95 \ {\rm km/h}$),  SV4 ($25 \ {\rm m}$, $95 \ {\rm km/h}$), EV ($0 \ {\rm m}$, $95 \ {\rm km/h}$). Similarly, SV3 will start to brake after time step 6 ($1.92 \ {\rm s}$) with a deceleration $-1.2 \ {\rm m/s^2}$, while the SV4 is controlled by the model \eqref{eq_ls_a}--\eqref{eq_ls_c} to start to change lane at time step 11 ($3.52 \ {\rm s}$), and its maneuver is predicted by the EV using the motion-prediction function of the IAIMM-KF. This makes the intention of SV4 unknown to the EV. The controller gains of SV4 are randomly sampled from the sets $\mathcal{K}_{\rm lon}^{m_5}$ and $\mathcal{K}_{\rm lat}^{m_5}$ at every time step, as SV4's maneuver corresponds to maneuver $m_5$ in Fig.~\ref{Fig_HighD_illustration}. This models the trajectory uncertainty of SV4 because the EV uses the nominal controller gains $\hat{K}_{\rm lon}^{m_5}$ and $\hat{K}_{\rm lat}^{m_5}$ to predict the trajectory of SV4, while the actual controller gains of SV4 are uncertain and time-varying. 

\begin{figure}[!t]
	\centering
	\subfloat[ ]{\includegraphics[width = \columnwidth]{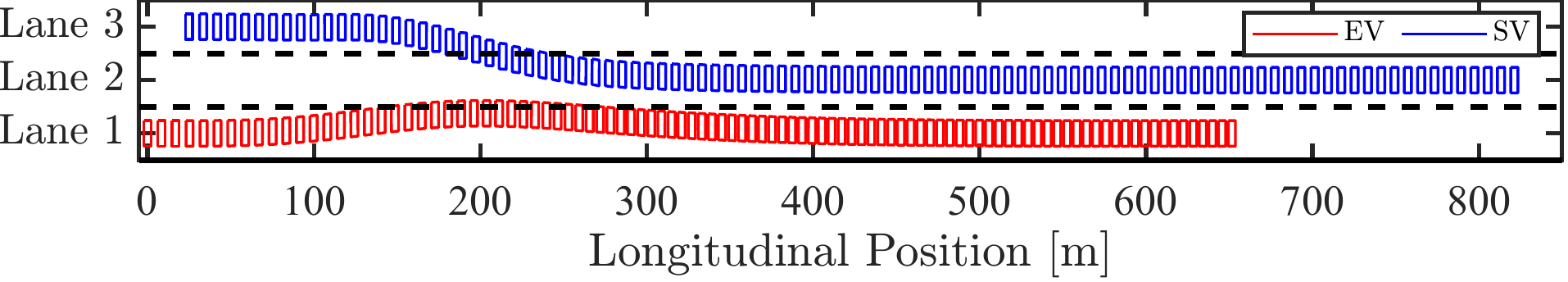}%
		\label{Fig_SIM_2_ev_sv_global_01}} 
	\vspace{0 cm}
	\subfloat[ ]{\includegraphics[width = \columnwidth]{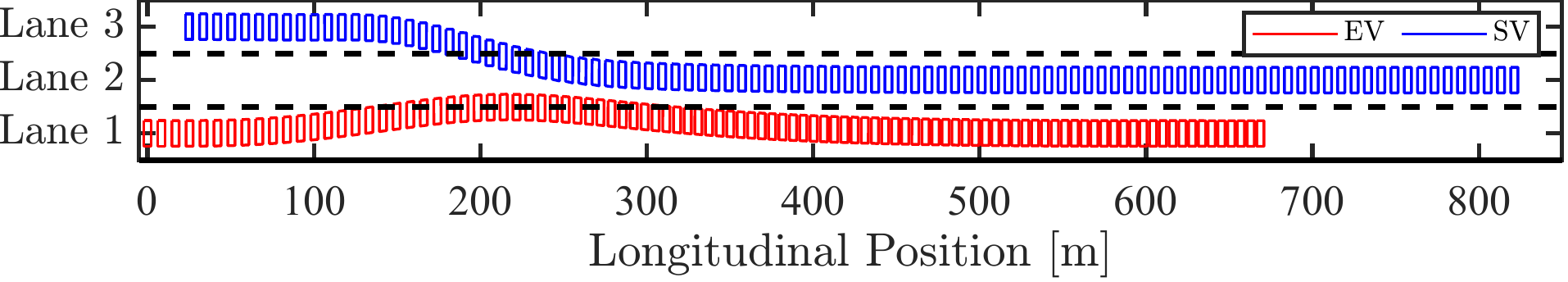}%
		\label{Fig_SIM_2_ev_sv_global_04}}
	\vspace{0 cm}
	\subfloat[ ]{\includegraphics[width = \columnwidth]{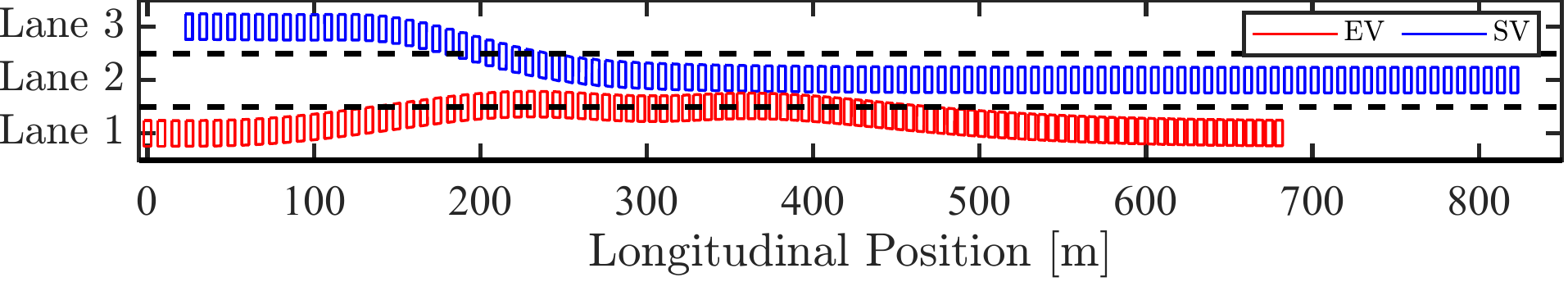}%
		\label{Fig_SIM_2_ev_sv_global_1}}
	\caption{Global paths by ISA-MPC and deterministic MPC. (a) Global path by ISA-MPC with $\varepsilon_k = 0.1$. (b) Global path by ISA-MPC with $\varepsilon_k = 0.4$. (c) Global path by the deterministic MPC.}
	\label{Fig_SIM_2_Global}
\end{figure}

The comparisons between the ISA-MPC and the deterministic MPC in one stochastic experiment are visualized in Fig.~\ref{Fig_SIM_2_Global}. It is seen that the deterministic MPC makes the EV closer to SV4. The ISA-MPC makes the distance between EV and SV4 larger, and a smaller $\varepsilon_k$ increases the distance. 

Since the maneuver of SV4 is random, it makes sense to evaluate the method from a statistical perspective. For this purpose, 300 Monte Carlo simulations were run for this scenario, and for each simulation, the minimum Euclidean distance (MED) between the center of EV and SV4 is recorded. The MED of the simulation with the deterministic MPC is represented as $\tilde{d}_{\rm min}$, and that with the ISA-MPC is denoted as $d_{\rm min}(\varepsilon_k)$, where $\varepsilon_k$ indicates the ISA-MPC with $\varepsilon_k=0.1$ and $\varepsilon_k=0.4$, respectively. 

\begin{figure}[!t]
	\centering
	\includegraphics[width = \columnwidth]{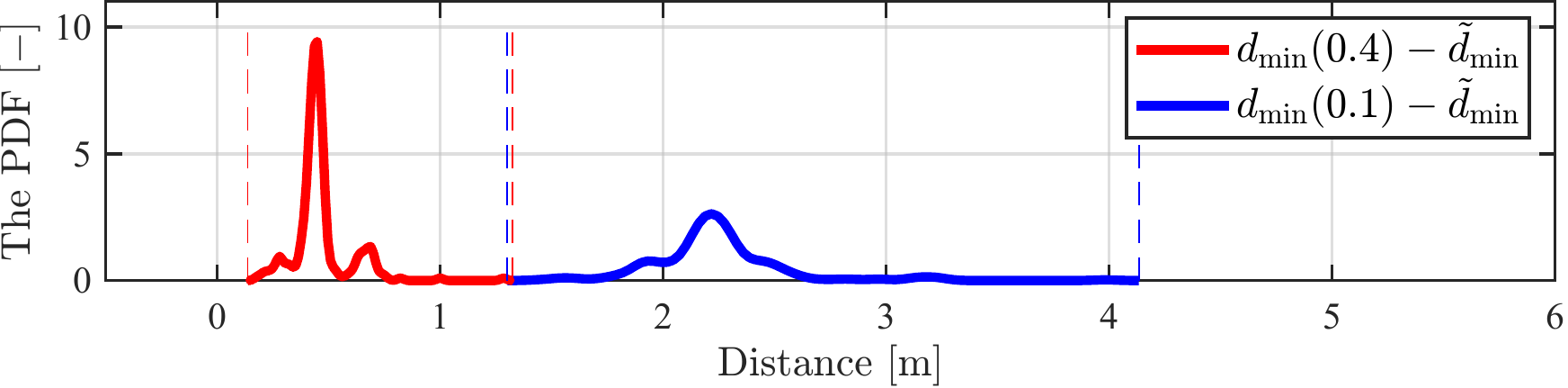}
	\caption{The estimated distributions of $d_{\rm min}(\varepsilon_k) - \tilde{d}_{\rm min}$.}
	\label{Fig_SIM_2_min_distance}
\end{figure}

The results of 300 simulations are used to estimate the distribution of $d_{\rm min}{(\varepsilon_k)}-\tilde{d}_{\rm min}$, as shown in Fig.~\ref{Fig_SIM_2_min_distance}. It is seen that the support of $d_{\rm min}(\varepsilon_k) - \tilde{d}_{\rm min}$ is always positive, which implies that in all conducted simulations, the MED between SV4 and the EV with the ISA-MPC is always larger than that with the deterministic MPC. Since a smaller MED indicates a lower safety level, we can conclude that the ISA-MPC is beneficial to increase driving safety in uncertain situations. In addition, the random variable $d_{\rm min}(0.1)-\tilde{d}_{\rm min}$ is distributed around the mean $2.21 \ {\rm m}$, while $d_{\rm min}(0.4) - \tilde{d}_{\rm min}$ is distributed around the mean $0.43 \ {\rm m}$. This means that reducing the value of $\varepsilon_k$ in the ISA-MPC increases driving safety. This is consistent with the expectations of the proposed approach.

\subsection{{\rm Case III}: Verification on the HighD Dataset}\label{SIM3}
\begin{figure}[!t]
	\centering
	\includegraphics[width = \columnwidth]{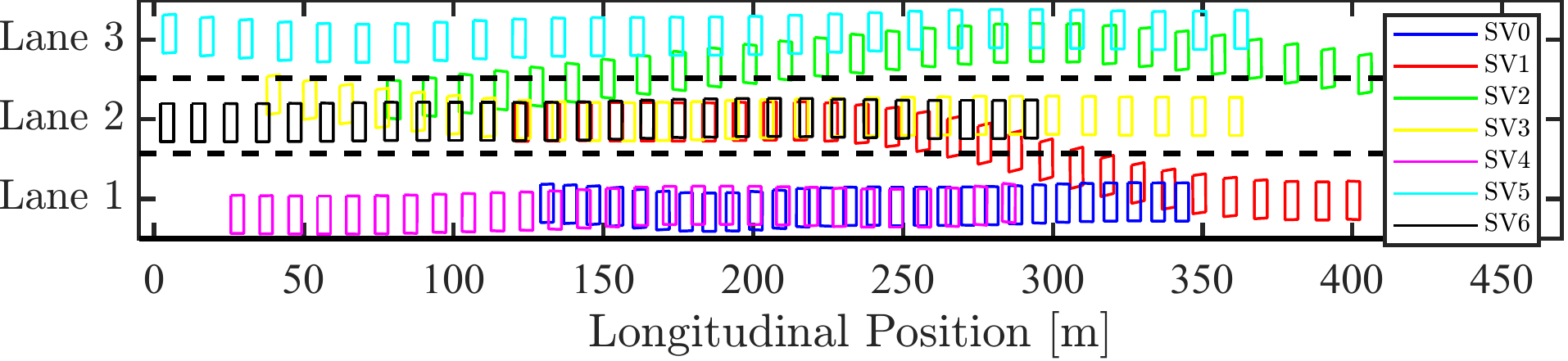}
	\caption{The highD dataset scenario for simulation in Section~\ref{SIM3}.}
	\label{Fig_SIM_HD_Used}
\end{figure}

In the third case, one scenario from the highD dataset \cite{krajewski2018highd} is taken to evaluate the performance of ISA-MPC compared with the original trajectory of a human-driven vehicle. The scenario is described in Fig.~\ref{Fig_SIM_HD_Used}, where seven vehicles are present at the recorded time period. Surrounding vehicles SV1 and SV2 perform lane-changing maneuvers and the others keep the lane. The initial longitudinal position, lateral position, and longitudinal velocity of the vehicles from the highD dataset scenario are: SV0 ($131.18 \ {\rm m}$, $1.67 \ {\rm m}$, $86.72 \ {\rm km/h}$), SV1 ($122.79 \ {\rm m}$, $5.49 \ {\rm m}$, $118.60 \ {\rm km/h}$),  SV2 ($80.05 \ {\rm m}$, $6.56 \ {\rm m}$, $134.14 \ {\rm km/h}$), SV3 ($39.81 \ {\rm m}$, $6.79 \ {\rm m}$, $137.02 \ {\rm km/h}$), SV4 ($27.82 \ {\rm m}$, $1.15 \ {\rm m}$, $108 \ {\rm km/h}$), SV5 ($5.09 \ {\rm m}$, $9.66 \ {\rm m}$, $145.15 \ {\rm km/h}$), SV6 ($4.37 \ {\rm m}$, $5.48 \ {\rm m}$, $120.56 \ {\rm km/h}$). Selected parameters from Table~\ref{Table_simulation_parameters} are tuned to fit the real traffic situation: $(W_x, W_y, W_v, W_l) =(0.2, 0.1, 0.5, 0.2)$, $\tau_h = 1$ s in \eqref{eq_mt_velocity_b}, and $\tau_h = 0.5$ s in \eqref{eq_spe_ocp_d}. The other parameters are consistent with those in Table~\ref{Table_simulation_parameters}.

\begin{figure}[!t]
	\centering
	\subfloat[ ]{\includegraphics[width = \columnwidth]{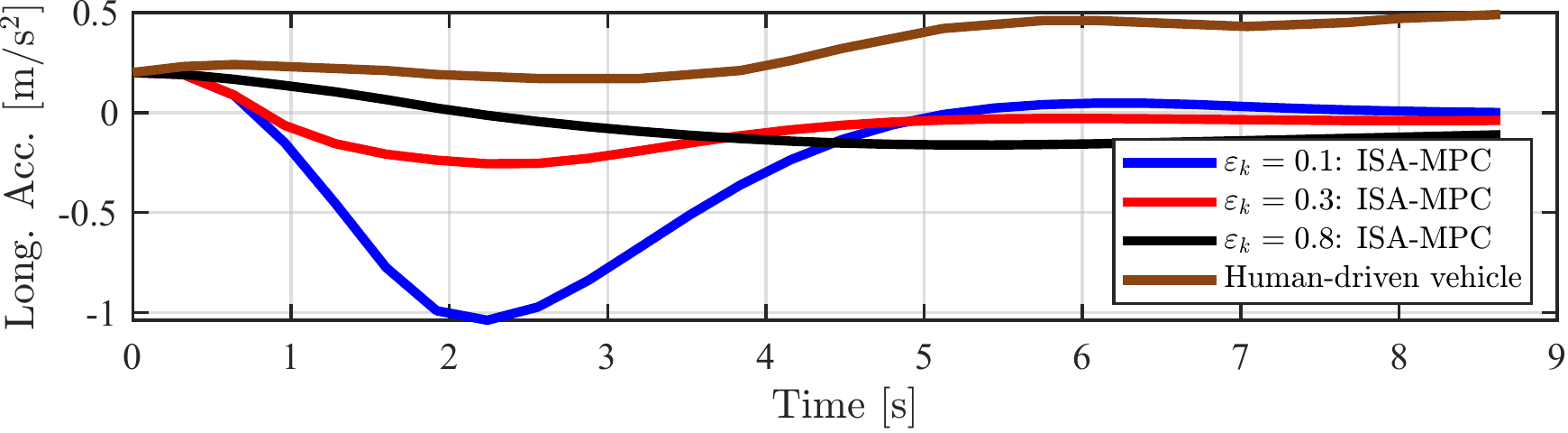}%
		\label{Fig_SIM_3_ax}}
	\vspace{0 cm}
	\subfloat[ ]{\includegraphics[width = \columnwidth]{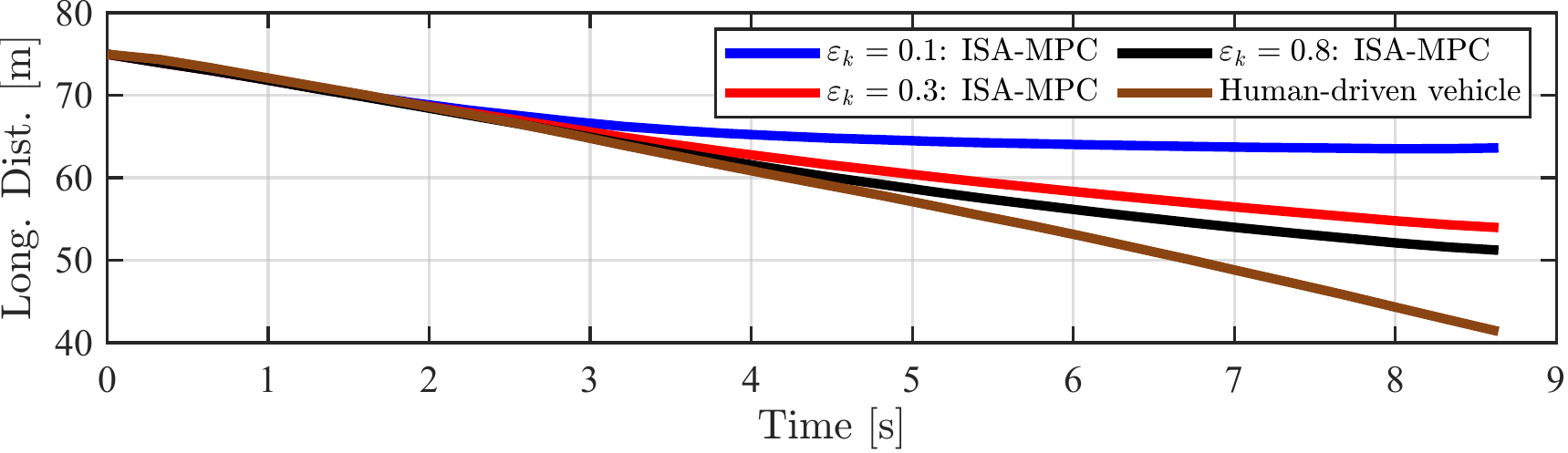}%
		\label{Fig_SIM_3_distance}}
	\caption{Comaprison between the ISA-MPC and the real vehicle from the highD dataset. (a) Longitudinal acceleration. (b) Longitudinal distance to SV2.}
	\label{Fig_SIM_3}
\end{figure}

To compare the ISA-MPC with the human-driven vehicle, a virtual vehicle is introduced in Fig.~\ref{Fig_SIM_HD_Used} as the EV, which has the same initial conditions as SV5. So, the trajectory of the virtual EV planned by ISA-MPC will be compared with the original trajectory of SV5. For the virtual EV, the SVs are predicted by IAIMM-KF, and the ISA-MPC is used to plan the motion with different safety-awareness parameters $\varepsilon_k$, and the results are shown in Fig.~\ref{Fig_SIM_3}. Since there is no motivation for SV5 and the virtual EV to change lanes in this scenario, both SV5 and the virtual EV will perform lane-keeping maneuvers, hence they are compared from the longitudinal perspective. 

Fig.~\ref{Fig_SIM_3}(a) shows the longitudinal accelerations planned by the ISA-MPC and the acceleration of the human-driven vehicle (SV5). It is seen that a smaller $\varepsilon_k$ results in a more wary longitudinal maneuver of the virtual EV, which tries to decelerate to increase the longitudinal distance to the obstacle SV2. Fig.~\ref{Fig_SIM_3}(b) visualizes the longitudinal distances between the EV and SV2 by the ISA-MPC and the human-driven vehicle. The ISA-MPC subjected to a safety-distance constraint is more conservative than the human driver, and the safety distance increases by reducing the safety-awareness parameter $\varepsilon_k$. 

\subsection{Computational Aspects and Future Work}\label{Discussion}
The simulations were performed on a standard laptop with an Intel i7-10750H CPU,
32.0 GB RAM running Ubuntu 22.04 LTS and Python 3. Table~\ref{Table_computation_time} presents the average computation time at every time step with the motion planner executed 1980 time steps in each case, where Planning Time means the computation time for running the whole motion planner, and MPC Solution Time means the time for solving the OCP \eqref{eq_spe_ocp_a}--\eqref{eq_spe_ocp_e}. It is seen that the method is real-time implementable in the simulations as all of the computations are finished within the sampling interval ($0.32$~s). Another observation is that changing the value of $\varepsilon_k$ of the ISA-MPC in the same traffic situation basically does not affect the computational complexity of both running the motion planner and solving the MPC problem, this is different from the SCMPC where the computation time is increased when increasing $K_{\rm sc}$. In Case III, it takes a longer time to run the algorithm as more SVs are involved in the scenario, while the MPC solution time generally does not increase with more SVs. 

In this paper, the value of the safety-awareness parameter $\varepsilon_k$ does not adapt to changes in the surrounding traffic. Hence, it would be interesting as future research to design a risk-aware method to adaptively compute the value of $\varepsilon_k$ according to the specific traffic situation. This can be done, e.g., by assessing the risk of collision with SVs\cite{parseh2021collision}.

\begin{table}[!t]\scriptsize
	\begin{threeparttable}
	\centering  
	\caption{Computation time for different cases}  
	\label{Table_computation_time}  
	\setlength{\tabcolsep}{1mm}
	\begin{tabular}{clcccc}  
		\toprule  
		\multicolumn{2}{c}{\underline{\textbf {Simulation Case}}} & \multicolumn{2}{c}{\underline{\textbf {Planning Time}}} & \multicolumn{2}{c}{\underline{\textbf {MPC Solution Time}}} \\
		\multicolumn{1}{c}{\textbf{Case Index}} & \multicolumn{1}{c}{\textbf{Planner}} & \multicolumn{1}{c}{\textbf{Mean}} & \multicolumn{1}{c}{\textbf{STD}} & \multicolumn{1}{c}{\textbf{Mean}} & \multicolumn{1}{c}{\textbf{STD}} \\
		\midrule
		${\rm I}$ & ISA-MPC ($\varepsilon_k = 0.2$) & 0.174 ${\rm s}$ & 0.013 ${\rm s}$ & 0.032 ${\rm s}$ & 0.003 ${\rm s}$ \\ \specialrule{0em}{1pt}{1pt}
		${\rm I}$ & ISA-MPC ($\varepsilon_k = 0.5$) & 0.178 ${\rm s}$ & 0.014 ${\rm s}$ & 0.034 ${\rm s}$ & 0.005 ${\rm s}$ \\ \specialrule{0em}{1pt}{1pt}
		${\rm I}$ & ISA-MPC ($\varepsilon_k = 0.8$) & 0.173 ${\rm s}$ & 0.013 ${\rm s}$ & 0.032 ${\rm s}$ & 0.004 ${\rm s}$\\ \specialrule{0em}{1pt}{1pt}
		${\rm I}$ & SCMPC ($K_{\rm sc} = 5$) & 0.115 ${\rm s}$ & 0.005 ${\rm s}$ & 0.030 ${\rm s}$ & 0.001 ${\rm s}$\\ \specialrule{0em}{1pt}{1pt}
		${\rm I}$ & SCMPC ($K_{\rm sc} = 10$)& 0.154 ${\rm s}$  & 0.007 ${\rm s}$& 0.033 ${\rm s}$ & 0.002 ${\rm s}$\\ \specialrule{0em}{1pt}{1pt}
		${\rm I}$ & SCMPC ($K_{\rm sc} = 15$) & 0.193 ${\rm s}$ & 0.010 ${\rm s}$ & 0.036 ${\rm s}$ & 0.004 ${\rm s}$\\ \specialrule{0em}{1pt}{1pt}
		${\rm II}$ & ISA-MPC ($\varepsilon_k = 0.1$) & 0.178 ${\rm s}$ & 0.011 ${\rm s}$  & 0.034 ${\rm s}$ & 0.004 ${\rm s}$\\ \specialrule{0em}{1pt}{1pt}
		${\rm II}$ & ISA-MPC ($\varepsilon_k = 0.4$) & 0.174 ${\rm s}$  & 0.008 ${\rm s}$ & 0.032 ${\rm s}$ & 0.002 ${\rm s}$\\ \specialrule{0em}{1pt}{1pt}
		${\rm II}$ & Deterministic MPC & 0.099 ${\rm s}$ & 0.004 ${\rm s}$ & 0.031 ${\rm s}$ & 0.002 ${\rm s}$\\ \specialrule{0em}{1pt}{1pt}
		${\rm III}$ & ISA-MPC ($\varepsilon_k = 0.1$) & 0.229 ${\rm s}$  & 0.033 ${\rm s}$ & 0.034 ${\rm s}$ & 0.004 ${\rm s}$\\ \specialrule{0em}{1pt}{1pt}
		${\rm III}$ & ISA-MPC ($\varepsilon_k = 0.3$) & 0.191 ${\rm s}$ & 0.026 ${\rm s}$ & 0.033 ${\rm s}$ & 0.005 ${\rm s}$\\ \specialrule{0em}{1pt}{1pt}
		${\rm III}$ & ISA-MPC ($\varepsilon_k = 0.8$) & 0.197 ${\rm s}$ & 0.032 ${\rm s}$ & 0.035 ${\rm s}$ & 0.005 ${\rm s}$\\ \specialrule{0em}{1pt}{0.1pt}
		\bottomrule
	\end{tabular}
	\begin{tablenotes}\tiny
		\item ${\star}$The computation time for motion prediction of SV5 is not included in Case III.
	\end{tablenotes} 
\end{threeparttable}
\end{table}

\section{Conclusions}\label{Conclusions}
An interaction-aware resilient motion-planning method for an autonomous ego vehicle in uncertain multi-vehicle scenarios has been developed. The method integrates the ability of IAIMM-KF to predict interactive multi-modal maneuvers of surrounding vehicles, and the MPC's advantage in planning an optimal trajectory in dynamic environments. The multi-modal motion uncertainties quantify maneuver and trajectory uncertainties of surrounding vehicles. The prediction uncertainties are fully considered in the collision-avoidance constraints of the MPC, where a safety-awareness parameter plays an important role in tuning the trade-off between the performance and robustness of the method. Simulations show that (i) the method that considers multi-modal uncertainties can increase driving safety in uncertain driving scenarios; (ii) the expected performance of the method can be well-tuned by changing the value of the safety-awareness parameter; (iii) the computational efficiency of the method is not affected by the value of the safety-awareness parameter.

\bibliographystyle{IEEEtran}
\bibliography{IEEEabrv, mybib}
 
\vspace{-30pt}
\begin{IEEEbiography}[{\includegraphics[width=1in,height=1.25in,clip,keepaspectratio]{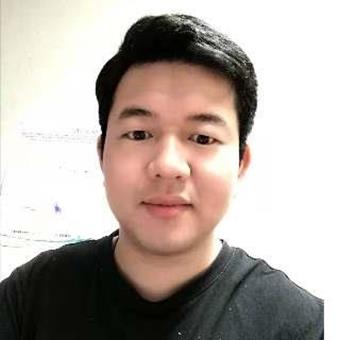}}]{Jian Zhou}
 received the B.E. degree in vehicle engineering from
the Harbin Institute of Technology, China, in
2017, and the M.E. degree in vehicle engineering from Jilin University, China, in 2020. He is currently a Ph.D. student with the Department of Electrical Engineering, Link\"oping University, Sweden. His
research interests are motion planning and control for autonomous vehicles and optimization with application to autonomous driving.
\end{IEEEbiography}
\vspace{-40pt}
\begin{IEEEbiography}[{\includegraphics[width=1in,height=1.25in,clip,keepaspectratio]{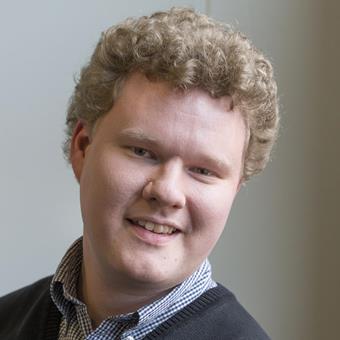}}]{Bj\"orn Olofsson}
	received the M.Sc. degree in Engineering Physics in 2010 and the Ph.D. degree
	in Automatic Control in 2015, both from Lund University, Sweden. He is currently an Associate Professor at the Department of Automatic Control, Lund University, Sweden, and also affiliated with the Department of Electrical Engineering, Link\"oping University, Sweden. His research includes motion control for robots and vehicles, optimal control, system identification, and statistical sensor fusion.
\end{IEEEbiography}
\vspace{-40pt}
\begin{IEEEbiography}[{\includegraphics[width=1in,height=1.25in,clip,keepaspectratio]{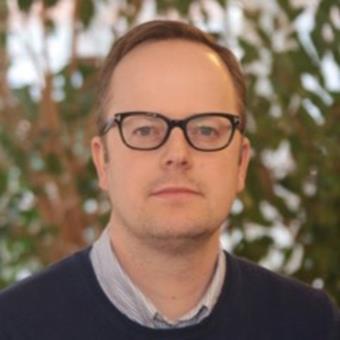}}]{Erik Frisk}
	was born in Stockholm, Sweden, in 1971. He received the Ph.D. degree in Electrical Engineering in 2001 from Link\"oping University, Sweden. He is currently a Professor with the Department of Electrical Engineering, Link\"oping University, Sweden. His main research interests are optimization techniques for autonomous vehicles in complex traffic scenarios and model and data-driven fault diagnostics and prognostics. 
\end{IEEEbiography}

\vfill

\end{document}